\documentclass[authoryear,12pt]{elsarticle}
\pdfoutput=1

\usepackage{color}
\usepackage{graphicx}
\usepackage{psfrag}
\usepackage{lineno}
\usepackage{epsfig}
\usepackage{subfigure}
\usepackage{amsmath}
\usepackage{amsfonts}
\usepackage{amssymb}
\usepackage{lscape}
\usepackage{bm}

\usepackage{paralist}
\usepackage{fancyhdr}
\begin{document}
\newcommand{\figurewidth}{6cm}

\begin{frontmatter}

\title{Spectral quantification of nonlinear behaviour of the nearshore seabed and correlations with potential forcings at Duck, N.C., U.S.A.}
\author[1]{Magar, V.\corref{cor1}}
\ead{vmagar@plymouth.ac.uk}

\author[2]{Lefranc, M.}
\ead{marc.lefranc@univ-lille1.fr}

\author[3]{Hoyle, R. B.}
\ead{R.Hoyle@surrey.ac.uk}

\author[1]{Reeve, D. E.}
\ead{dominic.reeve@plymouth.ac.uk}

\cortext[cor1]{Corresponding author}
\address[1]{Coastal Engineering Research Group, Marine Institute, School of Marine Science and Engineering, University of Plymouth, Drake Circus, Plymouth PL48AA. Tel: +44 (0)1752 586137, Fax: +44 (0)1752 232638}
\address[2]{Laboratoire de Physique des Lasers, Atomes, Mol\'ecules (PhLAM) and
Centre d' \'Etudes et de Recherches Lasers et Applications (CERLA),
{UFR} de Physique, Bat. P5,
Universit\'e des Sciences et Technologies de Lille,
F-59655 Villeneuve d'Ascq CEDEX, France}
\address[3]{Department of Mathematics, University of Surrey, Guildford, Surrey, GU2 7XH, UK}

\begin{abstract}
Local bathymetric quasi-periodic patterns of oscillation are identified from monthly profile surveys taken at two shore-perpendicular transects at the USACE field research facility in Duck, North Carolina, USA, spanning 24.5 years and covering the swash and surf zones. The chosen transects are the two furthest (north and south) from the pier located at the study site. Research at Duck has traditionally focused on one or more of these transects as the effects of the pier are least at these locations. The patterns are identified using singular spectrum analysis (SSA). Possible correlations with potential forcing mechanisms are discussed by  1) doing an SSA with same parameter settings to independently identify the quasi-periodic cycles embedded within three potentially linked sequences: monthly wave heights (MWH), monthly mean water levels (MWL) and the large scale atmospheric index known as the North Atlantic Oscillation (NAO)  and 2) comparing the patterns within MWH, MWL and NAO to the local bathymetric patterns. The results agree well with previous patterns identified using wavelets and confirm the highly nonstationary behaviour of beach levels at Duck; the discussion of potential correlations with hydrodynamic and atmospheric phenomena is a new contribution. The study is then extended to all measured bathymetric profiles, covering an area of 1100m (alongshore) by 440m (cross-shore), to 1) analyse linear correlations between the bathymetry and the potential forcings using multivariate empirical orthogonal functions (MEOF) and linear correlation analysis and 2) identify which collective quasi-periodic  bathymetric patterns are correlated with those within MWH, MWL or NAO, based on a (nonlinear)  multichannel singular spectrum analysis (MSSA). MEOF and linear correlation analysis are equivalent, with spatial distributions of the linear correlation coefficient between bathymetry and the potential forcing being the same as those of the relevant EOF. Regions of high correlations are supported by what would be expected from knowledge about local physical processes. At the two transects furthest from the pier, regions with high correlation and anti-correlation broadly agree with the SSA findings.  The MEOF/correlation analysis was used as a preliminary step towards the identification of collective spatio-temporal patterns between all bathymetric profiles and the potential forcings.  This final analysis, based on MSSA, showed that no collective interannual patterns of oscillations are present throughout the bathymetry and the three potential forcings - which was to be expected as SSA showed the highly localised nature of the interannual bathymetric patterns. Also,  annual and semi-annual cycles within the bathymetry are strongly correlated to the monthly wave height, in agreement with the SSA findings. Other collective intra-annual cycles besides the semi-annual were identified; they were all correlated to the NAO.
\end{abstract}

\end{frontmatter}

\section{Introduction}
\label{sec:intro}

Understanding the evolution of beaches in the long term (at yearly to decadal time scales) poses an important challenge to researchers. This is because it is unclear how the mechanisms forcing the morphology, such as hydrodynamic and sediment transport processes, affect beaches at such time scales \citep{SouthgateEtAl2003,DoddEtAl2003}, and also because datasets from which yearly to decadal patterns of behaviour may be extracted are at present scarce. Because of the threat of climate change, improved understanding and better predictive tools are crucial, both for researchers and end-users alike. 

The term 'coastal morphodynamics' has been coined to describe the investigation of changes in the shape (or morphology) of coastal features such as beaches, sandbanks and so on. There are several modelling approaches adopted in coastal morphodynamics, for instance sequential models, equilibrium models, equation-based or data-based models.  In long-term morphodynamical modelling data-based approaches tend to be preferred because of the difficulties in assessing how the different processes affect the dynamics in the long-term, as was mentioned above. Equilibrium-based models of sandbar systems assume sandbar migration towards a wave height dependent equilibrium location \citep[see, e.g.][]{PlantEtAl1999}. Sequential models either analyse single cross-shore profiles or incorporate longshore variability; in the former the aim is an analysis of cross-shore sediment transport, in particular the characterisation of cyclic patterns, while in the latter the aim consists of identifying morphological beach states that would characterise morphological sequences and thus help understand and classify nearshore bars in terms of their shape and dynamics \citep{WrightShort1984,LippmannHolman1990}. An important aspect of sandbar dynamics is the observed interactions between sandbars when several bars are present. At Duck, such interactions have been studied by \citet{LippmannEtAl1993}, who observed significant nonlinear behaviour of inner bars when outer bars were present.

There are two possible signal selection criteria used to analyse nearshore morphodynamics. One  is to analyse spatial patterns, for instance the evolution of the sandbar crests and troughs or the evolution of the shoreline, and the other is to find patterns optimizing a statistical measure, for instance patterns identified with empirical orthogonal functions (EOFs), Canonical Correlation Analysis (CCA) or wavelet analysis. The spatial pattern approach has been extremely popular and has contributed significantly to the understanding of sandbar dynamics \citep[see][and references therein]{LippmannHolman1990,LippmannEtAl1993,WijnbergTerwindt1995,PlantEtAl1999,WijnbergKroon2002,Rozynski2003,KuriyamaEtAl2008,PapeRuessink2008}. These studies focus on descriptions of bar migration, bar amplitude evolution and development of bar asymmetry. Using an optimal statistical measure as a pattern identification criterion has also been very popular; these methods are particularly useful for temporal or spatio-temporal pattern analysis. For instance, EOFs identify the linear modes of maximal variance, while CCA leads to linear modes that maximize the correlations between two time series. CCA has been used recently at Duck to analyse correlations between wave climate and bathymetric evolution at Duck, with reasonable success \citep{LarsonKraus1994,HorrilloCaraballo_Reeve2008}. \citet{ReeveEtAl2008}, on the other hand, used wavelets to analyse dominant temporal patterns of behaviour at a cross-shore transect in Duck. In some cases, it has been possible to link the spatial patterns to the patterns optimizing a statistical measure \citep[see, e.g.,][] {LippmannEtAl1993,WijnbergTerwindt1995}; in others the interpretation of the patterns optimizing a statistical measure may be less obvious, for instance for the 1-2 yearly temporal patterns embedded in the seabed dynamics at Duck, identified by \citet{ReeveEtAl2008}, in which case additional analyses may be necessary to assess their origin. 

We postulate that different bathymetric regions have different nonlinear behaviour depending on their cross-shore and longshore location, and that several of the observed patterns are well correlated with patterns embedded within potential forcing mechanisms such as the monthly wave patterns, monthly sea level patterns or the NAO. Hence, nonlinear patterns of behaviour of bed levels at Duck will be quantified using singular spectrum analysis (SSA) and spectral density analysis SDA to characterise quasi-periodic oscillations. Possible correlations between these patterns and those embedded within large-scale phenomena that might potentially be at the origin of the observed behaviour will be dicussed, and correlations between them will be confirmed by further multivariate EOF (MEOF) and multichannel singular spectrum analysis (MSSA) studies.

SSA and MSSA are two data-based, nonlinear methods that optimize modes of maximal variance. The methods are applied to time-lagged copies of bathymetric (and hydrodynamic or atmospheric index) time series and were chosen because they are good at identifying underlying temporal and spatio-temporal patterns of oscillation even when the data is short and noisy \citep{VautardGhil1989}, as is the case for most coastal morphodynamics datasets.  \citet{VautardGhil1989} have shown, for instance, that SSA is capable of describing the dynamics of paleoclimatic oscillations obtained from marine sediment core measurements that had as few as $184$ data points, as well as the statistically significant number of degrees of freedom. There are other measures of dimension aside from the number of degrees of freedom of a dynamical system, such as the correlation dimension, which is a measure of the dimensionality of the space occupied by a set of points. However, \citet{Grassberger1986} and \citet{VautardGhil1989} have showed that reliable correlation dimension computations for the same core measurements were not possible due to the data's shortness and noisiness. Indeed, it is well known that in general the number of data points needed for accurate computation of the correlation dimension would be much larger, more likely in the order of a thousand or larger \citep{GrassbergerProcaccia1983,Ruelle1990}. When analysing temporal patterns, however, it is important as well that the frequency of sampling is significantly larger than the frequency of the studied pattern, regardless of the size of the dataset \citep{GilmoreLefranc2002}.  For our purposes  SSA (and by extension MSSA) is a perfectly suitable technique  since our interest focuses on the quasi-oscillations underlying the dynamics and these oscillations are well captured by such techniques. 

SSA has some similarities with empirical orthogonal functions (EOF), first applied to analysing beach variation by \citet{WinantEtAl1975}. Both SSA and EOF, as well as MSSA and Complex EOF (CEOF), have very recently been exploited to analyse long-term sea level changes \citep{PapadopoulosTsimplis2006}, as well as behavioural patterns in coastal features such as shorelines \citep{RozynskiEtAl2001,SouthgateEtAl2003,Rozynski2005,MillerDean2007a} and nearshore sandbars \citep{WijnbergTerwindt1995,RattanEtAl2005}. SSA may be used together with spectral density analysis (SDA) to characterise the frequencies identified with SSA. However, one major benefit of SSA over spectral analysis is that, just as with wavelet analysis, SSA is capable of capturing nonstationarity in patterns \citep{LiEtAl2005,ReeveEtAl2007}.

\citet{SouthgateEtAl2003} used SSA to identify the long-term trend of the shoreline at Duck, between $1980$ and $1993$. The authors analysed  the component of the signal resolved by the first three eigenvalues. Their reconstruction showed no evidence of quasi-periodic oscillations of the shoreline. Rather, it led to the identification of an accretional trend from 1981 to 1990, followed by significant erosion from 1990 onwards. This contrasted with their findings at Ogata beach, Japan, where a clear linear trend of erosion, with a $5$-yearly cycle superimposed, could be identified.

As \citet{RozynskiEtAl2001} indicate, the specification of what is a signal and what is noise is crucial in SSA. These authors have used this technique to analyse shoreline evolution at Lubiatowo, Poland, between $1983$ and $1999$. Three long-term patterns were identified, with periods of $9$, $16$ or $32$ years, at different positions along the coastline. Interannual cycles of  $2$-$3$ and $3$-$4$ years were also found. Significant variability occurred at some positions, however. This study also pointed to the possible existence of chaotic behaviour as resolved by the second eigenvalue, which the authors linked to a response to extreme events. An association between the rhythmicity of the bed level variations and sand bar propagation was also postulated, supported by the existence of high correlations between shoreline positions and the inner bar crests \citep{PruszakEtAl1999}. The authors also explain some findings in terms of self-organisation.
While the implementation of EOF is relatively straightforward, with
SSA intermediate computations and tests are necessary to obtain a set
of uncorrelated reconstructed components and avoid
misinterpretations. However, SSA is superior to EOF because it can identify
nonlinear, temporal patterns underlying the dynamics, while EOF is a linear method which cannot identify nonlinear patterns at all. MSSA follows the
same principles as SSA, but is performed on a vector that contains variations at different positions rather that at a single location. Thus, MSSA is more likely to identify collective patterns of behaviour. MSSA is sometimes called multivariate extended empirical orthogonal function (MEEOF) analysis \citep{MoteEtAl2000}.

MSSA has been applied to Lubiatowo, to identify the collective patterns of behaviour of the shoreline \citep{Rozynski2005}. However, such patterns will necessarily appear as well in the SSA at individual positions, but possibly not everywhere as the dominant patterns of behaviour. At Lubiatowo, three collective quasi-periodic oscillations were identified, with periods of $7$ to $8$~years, $20$~years and several decades. The authors commented that the $7$ to $8$-yearly cycle might be forced by the North Atlantic Oscillation (NAO), which has an important influence in the Baltic Sea. Very recently, \citep{Rozynski2010} extended his analysis and found supporting evidence that the NAO may affect the bathymetry through a gentle coupling between the NAO, the hydrodynamics and the seabed.

As indicated by \citet{FalquesEtAl2000}, the spatial regularity of nearshore patterns implies that nearshore dynamics may be understood and, most importantly, described using simple physical mechanisms. That we need to look for mechanisms acting in the long term is particularly important.  Within the surf zone and at the shore, the occurrence of spatially rhythmic patterns has been attributed to the presence of low-frequency waves, such as infragravity waves \citep{BowenGuza1978,Bowen1980,HolmanBowen1982}. However, infragravity waves are responsible for the short term response of some beaches, not the interannual behaviour. At such large time scales the wave-related parameter that is of importance is the monthly averaged wave height. Note that we are not implying there ought to be a simple relationship between the waves and the bathymetry, whose response to forcings has been shown to be nonlinear in other studies \citep[][and references therein]{FalquesEtAl2000}. This nonlinearity could be reflected in the difference between amplitudes and phases, as well as in the characteristics of the regime changes observed in the corresponding time series. In the case of longshore parallel bars, the bar crests and troughs may cause redistribution of wave energy, variations in the wave breaking point and wave transformation processes (i.e. wave refraction, reflection and diffraction) which in turn, produce a radiation stress which is no longer in equilibrium with the set-up and set-down, hence creating a steady circulation \citep{FalquesEtAl2000}. Cellular automata models have been developed to analyse free evolution of the seabed \citep{CocoEtAl1999}, while nonlinear morphodynamic evolution models have permitted the analysis of difference in bed response depending on an initial perturbation \citep[see, e.g.,][]{KleinSchuttelaars2006}.
Aside from the monthly averaged wave heights and atmospheric phenomena such as NAO, it is likely that the tides and storm surges also have an effect on long term morphodynamics. Tides are fully predictable shallow water waves generated by gravitational forces between the Earth, the Sun and the Moon.   Storm surges change the water depth through pressure forces, wind stresses and wave set-up and set-down, and this in turn changes the water depth at the bar crests and troughs, and hence the radiation stresses mentioned above. Given that tides and storm surges may be combined in one single parameter, namely the mean water level, some of the patterns observed in the bathymetric evolution may potentially be correlated with patterns found within a water level time series.

In Sect.~\ref{sec:methods} below the case study data is presented and the SSA and SDA techniques are explained in more detail, together with the MEOF and the MSSA. The results are presented in Sect.~\ref{sec:fund-freq} and they are then discussed in Sect~\ref{sec:discussion}. Sensitivity to window length changes is discussed in Sect.~\ref{sec:sensitivity}. Comparisons with previous investigations are presented in Sect.~\ref{sec:comparison_previous_studies}. In Sect.~\ref{dis_sec:wave}, SSA and SDA of the monthly averaged wave heights at a nearshore wave gauge are performed and the patterns are compared to those of the bathymetry, together with similar analyses for the North Atlantic Oscillation (taken as an average over the whole area of study) to illustrate the possible correlations with large-scale phenomena. Sect.~\ref{dis_sec:forcingsSVDandMSSA} discusses linear correlations between the full set of bathymetric measurements and the monthly mean wave heights, monthly mean sea level and monthly NAO. Coherent spatio-temporal correlations identified via MSSA are also discussed.  Finally, Sect.~\ref{sec:conclusions} contains a summary and some concluding remarks. Websites are referred to as data or software sources, when appropriate, in accordance with copyright requirements.

\section{Summary of data and techniques}
\label{sec:methods}

\subsection[Nature of data]{The source and nature of the data under analysis}

Data were provided by the Field Research Facility (FRF), Field Data Collections and Analysis Branch, US Army Corps of Engineers (USACE), Duck, N. C., available from http://www.frf.usace.army.mil.  The site consists of an open straight beach characterized by relatively regular, shore parallel contours with a barred surf zone and a moderate slope; these characteristics change  close to a pier by the FRF facility that extends from the dune to a nominal water depth of approximately $7$~m. The site has been comprehensively described by previous authors \citep{HowdBirkemeier1987,LeeBirkemeier1993,LeeEtAl1998,NichollsEtAl1998} and has also been the subject of numerous other studies (see introduction). 

Measurements have been taken since June 1981 on a monthly basis over a series of $26$ shore perpendicular transects, generally extending from the dune to approximately $950$~m offshore, except for 4 lines, located at approximately 500 and 600 metres north and south of the pier, which have been surveyed every fortnight.  Table 1 shows the averaged alongshore locations of these 4 lines with their associated measurement errors as well as their transect number (FRF location code).  Note that the alongshore coordinate is assumed to be $y$ and the cross-shore coordinate is $x$. The baseline is a shore parallel line with its origin at the South-East corner of the FRF property.  On-going quality control and early publication of the surveys for those transects led to a significant number of studies based on them (see introduction). Moreover, the bathymetric evolution at these profiles is the least affected by the presence of the pier\citep{NichollsEtAl1998}. However, important differences in the behaviour north and south of the pier, in particular regarding the sandbar system, have been observed \citep{ShandEtAl1999}. 

Duck is an example of a net offshore bar migration (NOM) system, or a system with interannual cyclic sandbar behaviour \citep{ShandEtAl1999,RuessinkTerwindt2000}. NOM is a cyclic phenomenon consisting of three stages \citep{RuessinkKroon1994}:  bar generation near the shoreline (stage 1); systematic offshore migration of the bar across the surf zone during high energy periods (stage 2) and; disappearance of the bar in the outer surf zone (stage 3).  Sandbar formation and migration may also be triggered by previous profile geometries, with the degeneration of an outer bar often leading to formation of a new bar near the shoreline \citep{RuessinkTerwindt2000}. NOM systems are often alongshore coherent, but at Duck this coherence is somewhat broken by the presence of the pier. Profiles at Duck vary from unbarred to triple barred, with the most common profile configuration being a double bar system consisting of a narrow inner bar and a broad outer bar \citep{HowdBirkemeier1987,LeeBirkemeier1993,LeeEtAl1998}.

Three parameters may characterise a NOM cycle at each stage: the average cross-shore distance over which bars migrate; the average duration of the bar migration and; the average return period of migration cycles. These parameters are site specific and in Duck they vary North and South of the pier. \citet[][Fig. 7]{ShandEtAl1999} summarise site parameter characteristics for a number of NOM systems including Duck North (DN) and Duck South (DS). While the total distance and the total duration of bar migration are relatively similar at DN and DS, the average return period is significantly different. The total distance travelled by the bars is 289.6 and 289.1 metres and their total duration (adding all three stages) is of 4.4 years and 4.1 years at DS and DN, respectively.
However, for stage 2 for instance, the most active stage, the return period at DS is 3.2 years while at DN it is 6.8 years. The sand is transported back to the shoreline during fairweather periods through a variety of mechanisms: in the outer surf zone transport is driven by wave asymmetry while in the inner surf zone it also occurs through migrating bodies such as transverse bars or bar bifurcates generated during  bar splitting processes \citep{HolmanSallenger1986,HolmanLippmann1987,LeeEtAl1998,Shand2007}. Since the wave climate has a very strong seasonal pattern with storms occurring mostly during the winter, annual patterns are expected to appear within the bathymetry. 

In this study, the two outermost transects, i.e. transects 58 (Duck North) and 190 (Duck South), were first chosen to analyse the local quasi-oscillations embedded in the most active part of the surf zone. This reduced the cross-shore range to $390$~m, between $x=100$ and $x=490$; this range coincided with that analysed by \citet{SouthgateMoller2000} and includes most of the NOM width, that is, the region of sandbar migration \citep[see][Fig. 3]{LeeEtAl1998}.  However the study area was extended to $x=80$~m (to include the shoreline processes in more detail) and to $x=520$~m (to check the type and strength of the correlations further offshore) for the linear correlation and MSSA analyses.] A spatial and temporal Akima interpolation \citep{Akima1970} was performed on the data to obtain depth values at constant $10$~m intervals and at even timesteps of $30$ days. These time and space interpolation intervals were also chosen by \cite{SouthgateMoller2000} and prevent over-interpolation of the data. This procedure led to a 299-point time series at every bathymetric position, spanning from July $1981$ to January $2006$. Cross-shore bathymetric anomalies (with respect to the time-average at each position along the transect) are shown in Figs (a,i) and (b,i) within Fig. \ref{fig:detrending}  for transects 58 and 190, respectively.



\subsection{Methods used for local pattern analysis}
Two complementary methods were used to identify local patterns of behaviour:  Singular Spectrum Analysis (SSA) and Spectral Density Analysis (SDA).  Both methods are described in detail in Appendix \ref{method_subsec:ssa_sda} so here we only present a brief summary of SSA. Essentially, SSA is a nonlinear eigenvalue problem for a M-dimensional sequence $Z_n=\left( z_n, z_{n+1}, \ldots, z_{n+M-1} \right)$, for $n=1,2, \ldots, N-M+1$, of time-lagged copies of the bed level $z(t)$, at times $t=i \tau_s$, $i=1,2, \ldots,N$, $N$ being the length of the time series, $\tau_s$ the sampling interval and $M<N$ is the window length. Time scales of the dynamics that can be reconstructed from this time series are between $\tau_s$ and the window time span. Reconstructed components (RCs), linked to the eigenvalues of $Z_n$, have the same time span as the original time series and isolate one or more or the quasi-oscillatory patterns (depending on the number of eigenvalues used for the reconstruction). In this way, the signal was separated into a trend and a detrended signal using a window length, $M$, of $4$~years. The trend corresponds to linear variations as well as patterns with periods above 3 to 5 years and is the RC of the first few eigenvalues of the SSA with $M=4$. The trend can be analysed further with an SSA with window length of, say, $9$~years, to identify long-period oscillations (LPOs) embedded within the trend signal. The detrended signal, on the other hand, contains short-period oscillations (SPOs), of periodicities of less than 5 years. Again, the periodicities of the SPOs can be determined with a further SSA on the detrended part of the signal. Note the overlap between LPOs and SPOs; this is due to the chosen window length of 4 years for the initial SSA.  The robustness of the observations to small changes in the window length will be discussed in more detail further on.

Patterns embedded within monthly mean wave heights (MWH), monthly mean water levels (MWL) and the monthly North Atlantic Oscillation (NAO) were also extracted following the same procedure. The details of the hydrodynamic and atmospheric data are discussed in Appendix  \ref{method_subsec:ssa_sda}. The patterns obtained were compared to those within the bathymetry to identify locations where similar patterns are observed. This led to an evaluation of correlations between bathymetric and hydrodynamic/atmospheric patterns. Note that SSA would give a nonlinear measure of the correlations. Linear correlations were found as a preliminary step for in the analysis of spatio-temporal patterns of behaviour; the correlation analysis and the methods involved in the spatio-temporal study are introduced in the next section. 

\subsection{Correlation analysis and spatio-temporal patterns between bathymetry and potential forcings}

Spatio-temporal collective bedlevel patterns were then analysed, not only on Transects 58 and 190 as before but also for all bathymetric surveys between them. A linear correlation analysis was first performed, by computing the correlation coefficient matrix, $r$, between the bedlevel time series and that of each of the potential forcings. This led to three different correlation coefficient matrices, one for the correlations of the bathymetry with the NAO, one for correlations with the MWH and one for correlations with the MWL, respectively. The values in $r$ can be used to produce correlation coefficient contours to identify locations where the correlation between the bedlevels and the forcing is strongest.  This  analysis is intrinsically linked with the MEOF analysis carried out as a preliminary step for MSSA for all the bathymetry transects, because the correlation coefficient is equivalent to the covariance matrix, $C$, normalised by the square root of the product of the covariance at the associated diagonal elements:
$$
r(i,j)=\frac{C(i,j)}{\sqrt{C(i,i) C(j,j)}},
$$
and MEOF is based on the covariance of the multivariate matrix. The MEOF method is described in detail in Sect.~\ref{subsec:meof_description} within the Appendix.
MEOF permits  reduction in the number of channels from 1128 (one channel for each bathymetric location and one channel for each of the three potential forcings evaluated at a single location - see the Appendix for details) to three channels corresponding to the MEOF PCs in directions of maximal variance of the potential forcings (see Table \ref{table_Ucomponents}), with EOF1 corresponding to NAO, EOF2 to the mean wave heights (MWH) and EOF3 to the mean water levels (MWL).  This reduction from 1128 to 3 channels is called a prefiltering. The system of 3 channels is then analysed with MSSA to identify which collective, bedlevel spatio-temporal patterns may be correlated with  NAO, MWH or MWL. The MSSA technique is described in Sec~\ref{subsec:meeof_description} in the  Appendix.

\section{Results}
\label{sec:fund-freq}

\subsection{Identification of trend and detrended signals} \label{sec:trend_signal}

\begin{figure}[ht]
\epsfig{file=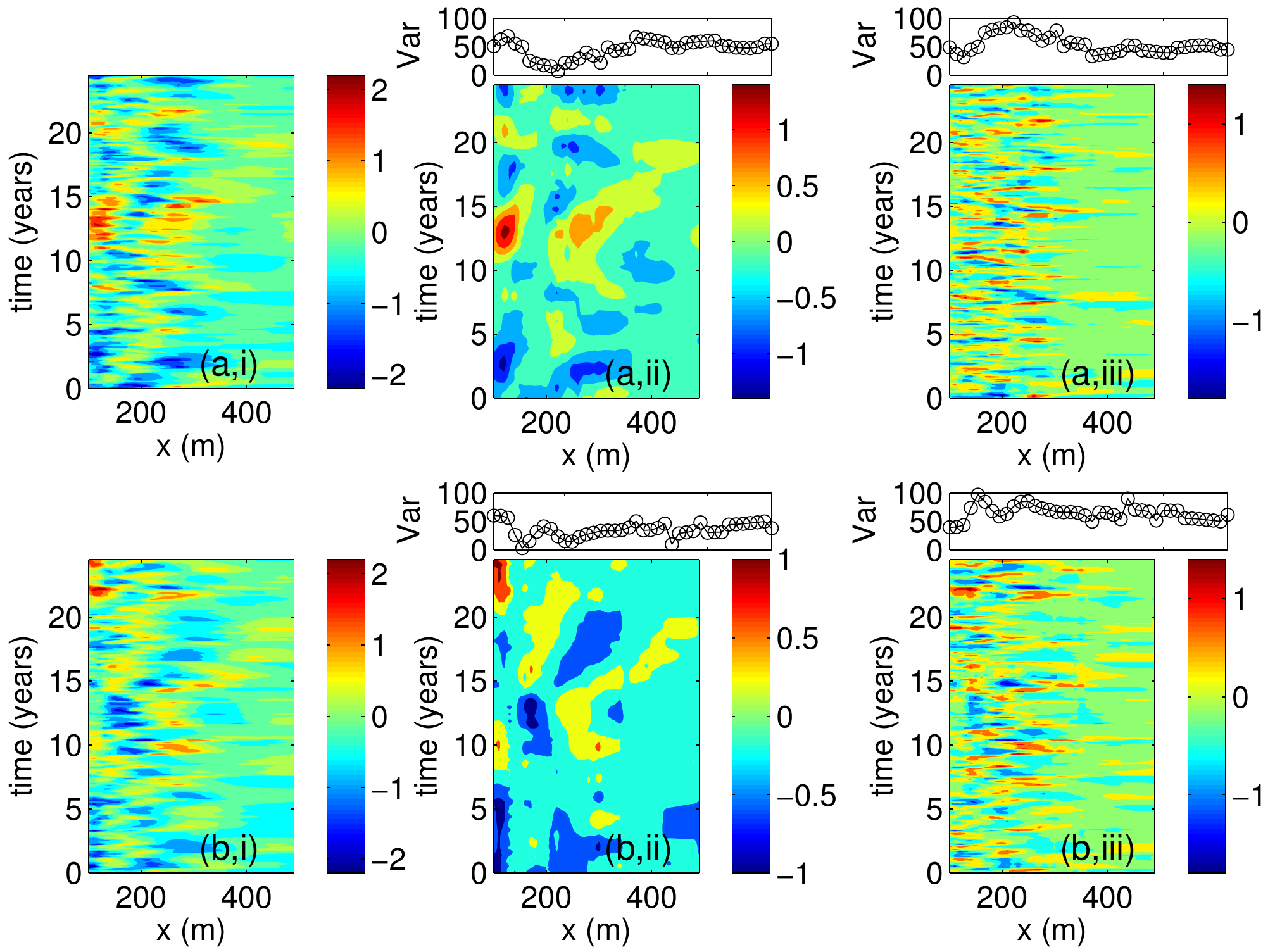,width=12.3cm}
\caption{\label{fig:detrending} Data contours for transects 58 (top, North of the pier) and 190 (bottom, South of the pier) with: beach elevation (in metres) displayed in Figs.~(a,i) and (b,i); trend signal elevation (in metres) displayed in Figs.~(a,ii) and (b,ii) and; detrended signal elevation (in meters) displayed in Figs. (a,iii) and (b,iii). The upper panels in Figs.~(a,ii), (a,iii), (b,ii) and (b,iii) show
  the percentage of variance resolved by the corresponding signal.}
\end{figure}

 
The trends, shown in Figs.~\ref{fig:detrending}(a,ii) and ~\ref{fig:detrending}(b,ii) for transects $58$ and $190$, respectively, capture well the dominant patterns observed in Figs.~\ref{fig:detrending}(a,i) and~\ref{fig:detrending}(b,i), which seem to correspond to the extrema in the bathymetry and therefore are likely to correspond to the sandbars extrema.  This is supported by the fact that some of the travelling characteristics of the sandbars have also been captured, because the extrema as well as surrounding bathymetric contours seem to travel offshore as time progresses.

For Transect $58$, the patterns extracted resolve approximately $50$\%
of the variance in regions where the extrema are well resolved. At
other locations, such as between $130$ and $220$~m offshore, the
variance resolved by the trend is slightly smaller than
$50$\%. However, these results indicate that the trend and the detrended signal are equally important because they resolve an equal proportion of the total variance, and hence both should be analysed further.

For Transect $190$, the results are very similar, i.e. the variance
resolved by the trend and the detrended signals is about the same,
though with the trend accounting for slightly less than $50$\% of the
variance in this case, and so both signals need to be analysed. As with transect $58$, the detrended signal, which contains the higher frequency oscillations (with periods generally below 4 years), seems to resolve slightly more of the variance between $130$ and $220$~m offshore.

The periods of the underlying cycle were computed all along transects $58$ and $190$; Table~\ref{table_peaks} shows the results at selected locations along the profiles. The uncertainties in the periods were estimated from figures of the power spectrum as well as from the uncertainty of the power values at each frequency. An example of a power spectrum is shown in Fig.~\ref{fig:density_plot_line58_x160_welch_80_4}, corresponding to position $x=160$~m at transect~58. If the uncertainty interval of the power value extends to or beyond the peak power value, then the uncertainty of the position of the peak also extends to the frequency at that power value.

\subsection{Identification of underlying quasi-oscillations} 
\label{subsec:quasiosc}

\begin{figure}[ht]
\centering
\subfigure[\label{fig:plate_plots_trend_QOs_W110_res_58} LPOs, Tr.~58]{\includegraphics[trim= 0 0 0 0,clip,width=5cm]{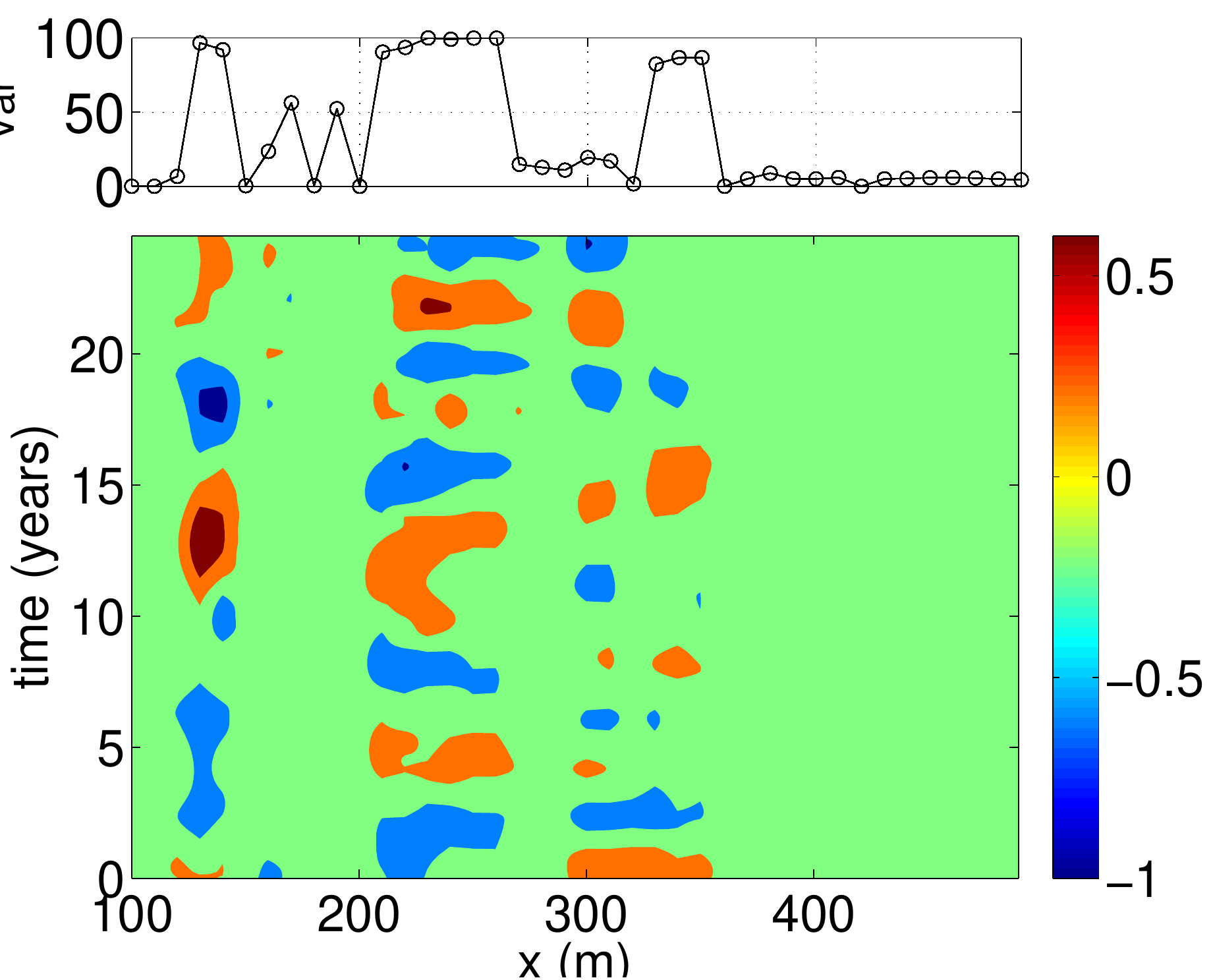}}
\subfigure[\label{fig:plate_plots_detrended_QOs_W110_res_58} SPOs, Tr.~58]{\includegraphics[trim= 0 0 0 0,clip,width=5cm]{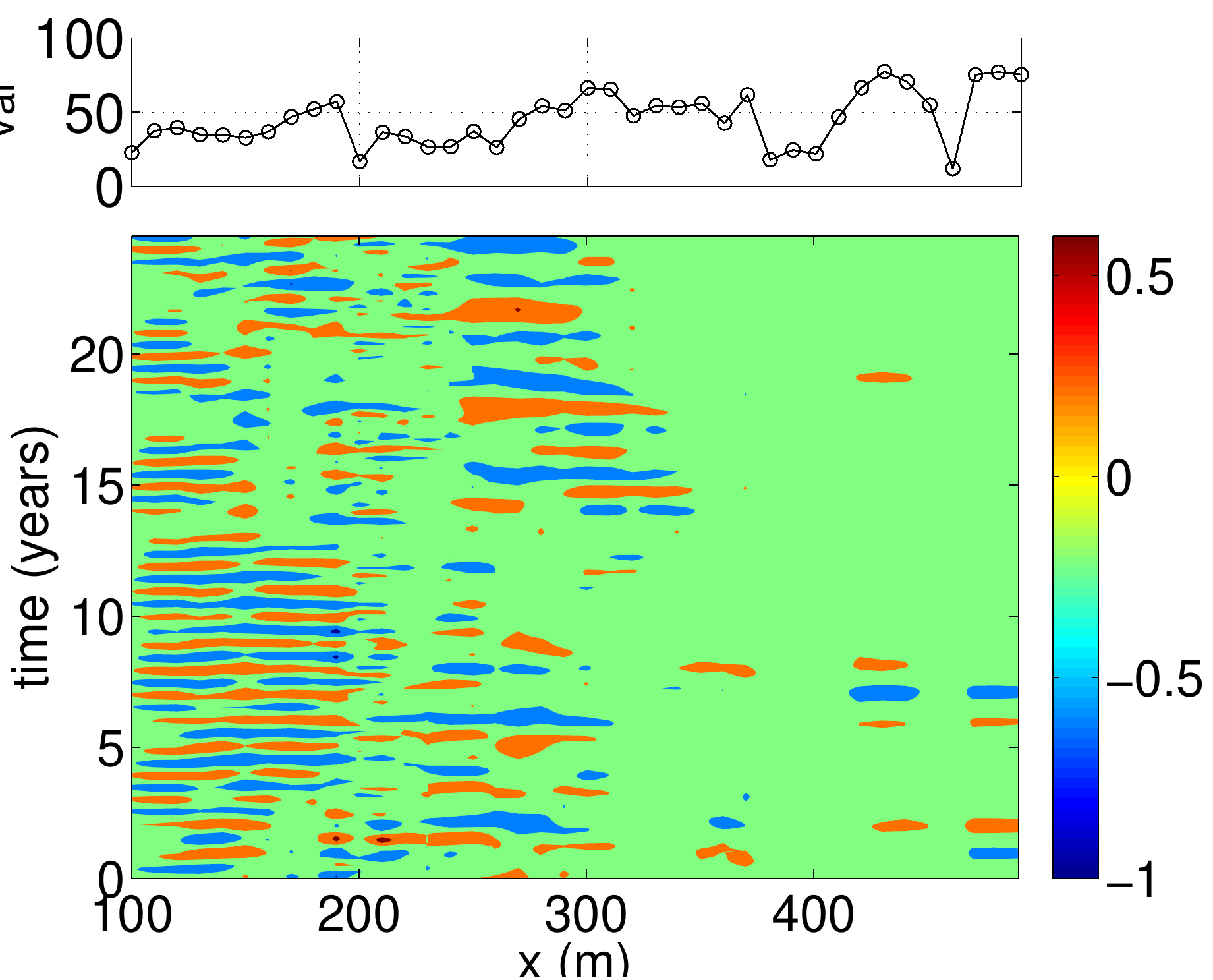}}\\
\subfigure[\label{fig:plate_plots_trend_QOs_W110_res_190} LPOs, Tr.~190]{\includegraphics[trim= 0 0 0 0,clip,width=5cm]{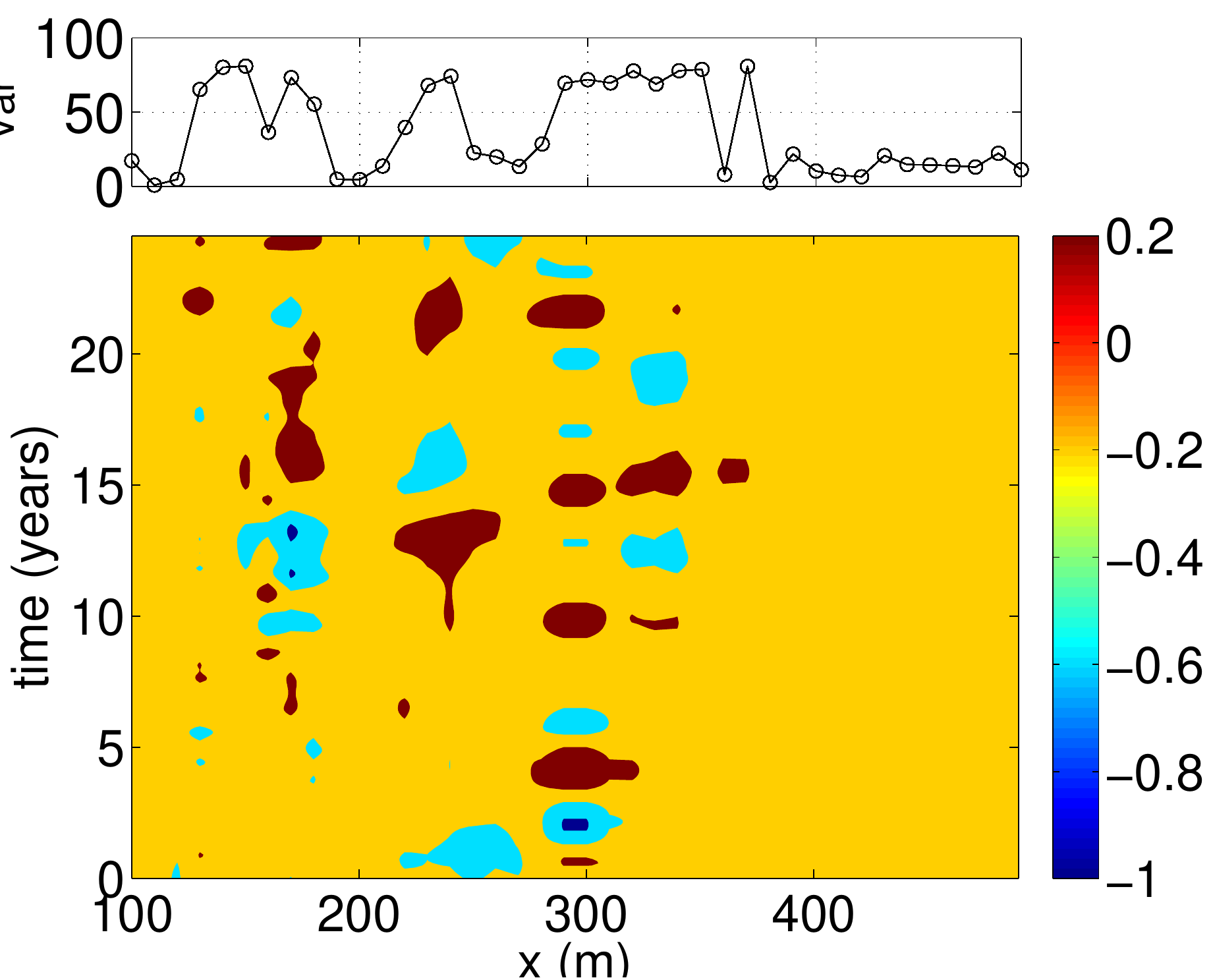}}
\subfigure[\label{fig:plate_plots_detrended_QOs_W110_res_190} SPOs, Tr.~190]{\includegraphics [trim= 0 0 0 0,clip,width=5cm]{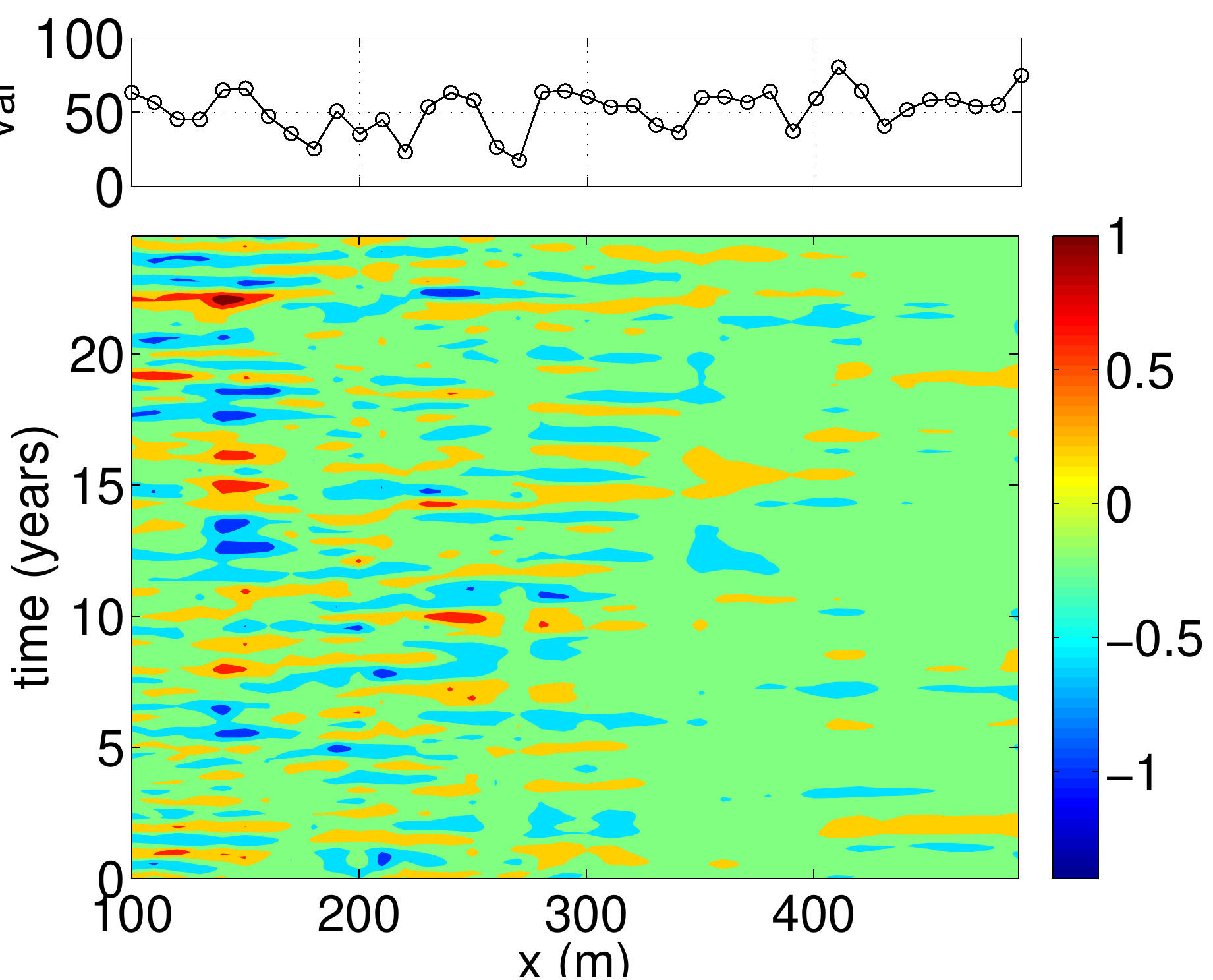}}
\caption{\label{fig:quasioscillations@110} Quasi-oscillations identified within the trend (left) and the detrended (right) signals together with the resolved variance, at window lengths of $9$~years, for transects $58$ (top) and $190$ (bottom).}
\end{figure}

\begin{figure}[ht]
\centering
\epsfig{file=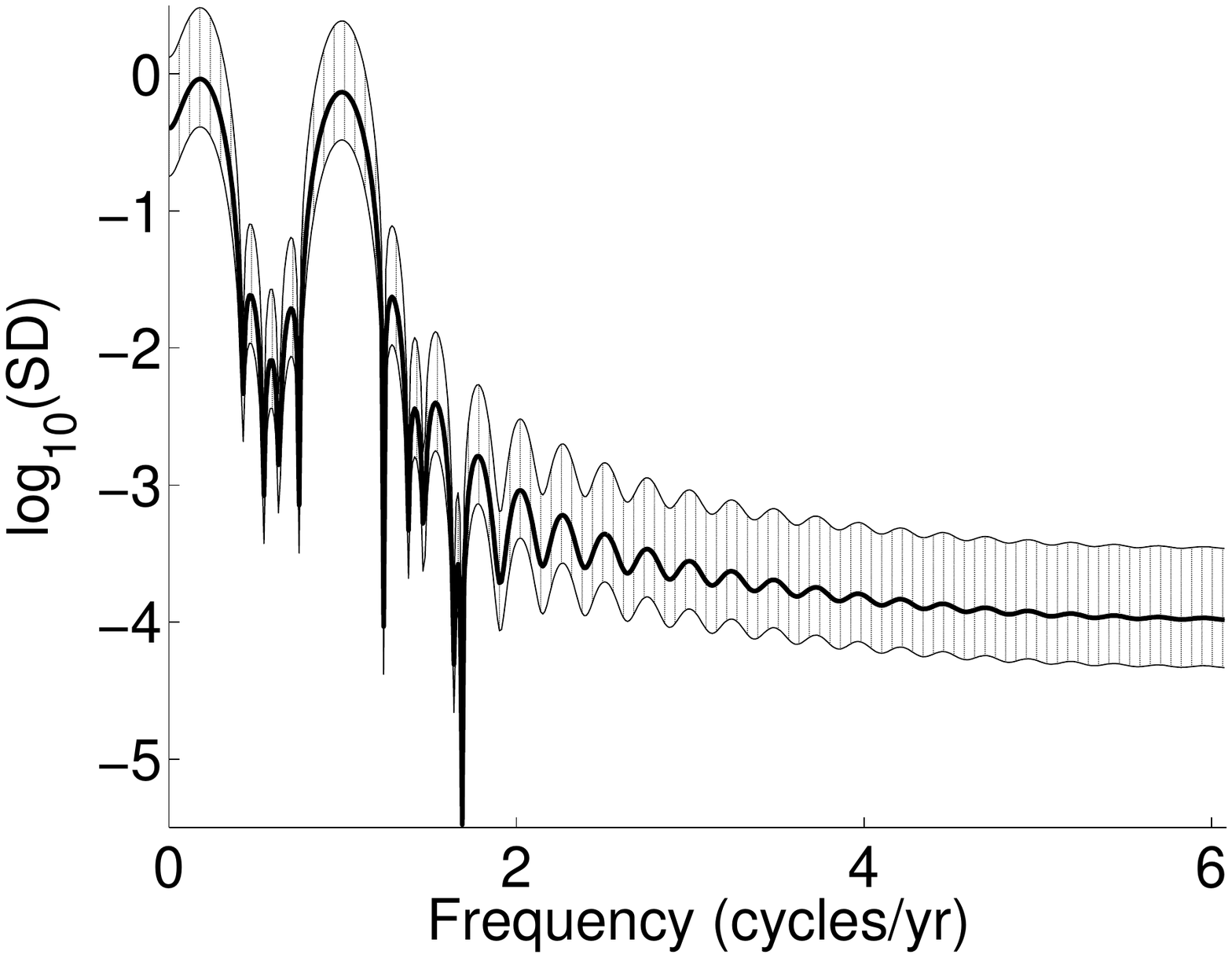,width=10cm}
\caption{\label{fig:density_plot_line58_x160_welch_80_4} Spectral density plot for the reconstructed component of the SPOs at $x=160$~m for Transect~$58$. The thin lines indicate the $95\%$ confidence interval for the spectral estimates.}
\end{figure}

Long period oscillations (LPOs) and short period oscillations (SPOs) - with periods above and below 4 years, respectively - were then extracted from the trend and the detrended time series. The RCs obtained when combining all the LPOs are shown in Figs.~\ref{fig:plate_plots_trend_QOs_W110_res_58} and~\ref{fig:plate_plots_trend_QOs_W110_res_190} for transects 58 and 190, respectively while the SPOs are shown in Figs.~\ref{fig:plate_plots_detrended_QOs_W110_res_58} and~\ref{fig:plate_plots_detrended_QOs_W110_res_190}.

In relation to the LPOs, it is clear that these patterns have an important contribution below $350$ to $400$~m offshore, because the variance resolved by these patterns - shown in the upper plot of Figs. ~\ref{fig:plate_plots_trend_QOs_W110_res_58} and~\ref{fig:plate_plots_trend_QOs_W110_res_190} -  typically takes large values there. At both transects the signal contains roughly three regions where quasi-oscillations are important. At transect $58$ these regions extend between $x=130$ and $140$~m, $x=210$ and $260$~m and $x=340$ and $350$~m; at transect $190$, they extend between $x=130$ and $180$~m, $x=230$ and $240$~m and $x=290$ to $370$~m. Hence there is agreement in the number of regions but their extent differs. The main periods of the quasi-oscillatory patterns are summarised in Table~\ref{table_peaks}. The table shows there is a lot of variability in the periods of the patterns, however, between $220$ and $370$~m offshore the periods of the patterns are close to those found within the NAO. 

The SPOs, in contrast, may have an important contribution throughout the region under consideration, as the variance they resolve - see upper plots in Figs.~\ref{fig:plate_plots_detrended_QOs_W110_res_58} and~\ref{fig:plate_plots_detrended_QOs_W110_res_190} - is usually around 50\%. Based on the information in Fig.~\ref{fig:plate_plots_detrended_QOs_W110_res_58} and
Table~\ref{table_peaks}, at transect $58$ the signal contains a very clear yearly 
cycle in the region between $x=100$ and $300$~m. A 1 to 2-yearly cycle is also present between $200$ and $300$~m offshore. From $x=300$ m onwards, the SPOs
have periods between $1.1$ and $1.3$~years around $x=310$ and between $2.2$ to $2.5$~years further offshore.
Beyond $300$~m, the amplitudes of the patterns
are very small, possibly because at these depths the effects of mechanisms driving such patterns are also smaller.  At transect $190$, examination of Fig.~\ref{fig:plate_plots_detrended_QOs_W110_res_190} and Table~\ref{table_peaks} shows the yearly cycle is present throughout the transect, together with a cycle whose period varies between two and three years, and from $200$~m to $490$~m sometimes with a third cycle with period between one and two years. The periods of the SPOs are close to those found within monthly wave heights, monthly sea water levels and again, within NAO, as we will discuss in subsequent sections.
 
It is noteworthy that between $350$ and $490$~m offshore the dominant
patterns have periods between $1$ and $9.8$ years at both transects, and that
interannual patterns of variable periods were also found closer to the
shore.  This suggests that differences in the temporal
patterns of the sandbar life cycle between different sides of the
pier are not reflected in differences of the seabed elevation variations between those sides.

Finally, at some locations the detrending seems not to have worked properly. This is the case, for instance, between $370$ and $390$~m for transect~$58$, or at $140$ and $200$~m for transect~$190$. In all cases the patterns have periodicities close to $4$ years and seem to have been misclassified within the trend or detrended signal. In some cases the uncertainty interval includes the window length size of $4$ years even though the value of the peak lies incorrectly in the trend or detrended signal. In other cases it is the lack of a pattern of period above $4$ years which leads to the identification of patterns between $3$ to $4$ years within the trend. Such issues are related to some limitations of SSA in identifying unequivocally patterns which have periodicities close to the size of the window length.

Patterns embedded within monthly mean wave heights (MWH), monthly mean water levels (MWL) and the monthly North Atlantic Oscillation (NAO) were also extracted following the same procedure and compared to those within the bathymetry to identify locations where similar patterns are observed. This will be discussed in more detail in Sect.~\ref{dis_sec:wave}. Linear correlations may also be computed; this was done throughout all the bathymetric surveys and the results are discussed next.

\subsection[Correlations between bathymetry and potential forcings]{Spatial and spatio-temporal correlations between the bathymetry, waves, mean water level and NAO}\label{res_sec:meof_mssa}

The questions we want to address in this section are whether there are correlations between the bathymetry and the NAO, the MWL and the MWH and whether these phenomena have a local influence or act throughout the whole bathymetry. These phenomena are evaluated at a single location, as explained in Sec.~\ref{sec:methods}, while there are measurements of the bathymetry at several locations along each transect used in our analyses.  The correlations between the bathymetry and the NAO, the MWH and the MWL are computed using a simple correlation analysis. The correlations obtained at all bathymetric locations are shown in Fig.~\ref{fig:Correlation_Contours}. The linear correlation analysis performed at all transects seems to corroborate the local correlations between embedded quasi-oscillations that were speculated about at transects 58 and 190 when comparing the LPOs and SPOs (shown in Table 2) to the atmospheric/hydrodynamic patterns deduced with SSA (see Sect.\ref{dis_sec:wave}). However, it is important to point out that a linear correlation analysis would permit neither to characterise the bedlevel patterns nor to suggest nonlinear correlations as was achieved with SSA.

The strength of the correlation is given by the correlation coefficient, $r$, shown in contour form in Figs~\ref{fig:Correlation_Contours}.  The maxima and minima of $r$ show locations where the correlations are strongest, independently of the magnitude of $r$. In order to assess the overall strength  of the (linear) correlation for each of the variables the following criteria may be useful: a) strong correlation when $0.5< \left\vert  r \right\vert \leqslant 1$; b) moderate correlation when $0.3<\left\vert  r \right\vert \leqslant  0.5$; c) weak correlation when $0.1<\left\vert  r \right\vert \leqslant  0.3$ and; d) no or very weak correlation when $0 <\left\vert  r \right\vert \leqslant  0.1$. With these criteria, the percentage correlation between bedlevels and potential forcings may be computed for all the bathymetric profiles.  If all the bathymetric surveys are considered, the overall strength of the linear correlations are as follows: 65.6\% of the bathymetry is very weakly correlated and 34.4\% is weakly correlated with NAO; 67.6\% of the bathymetry is very weakly correlated and 32.4\% is weakly correlated with MWH and; 28.3\%of the bathymetry is very weakly correlated, 50.5\% is weakly correlated and 21.2\% is moderately correlated with MWL. Note the values of these correlation coefficients are very low, but they only give a measure of the linear correlation between the bathymetric evolution and the potential forcings, while SSA and MSSA give a nonlinear measure of correlations. 

Table~3 shows for the first 4 EOFs from the MEOF analysis, the rows corresponding to each of the potential forcings, for transects $58$ and $190$ and then for the full set of bathymetric measurements at locations between and including these two transects.  The potential forcings have been organised so that the one at the top has the largest (absolute value) contribution for EOF1, the second for EOF2, and so on. It is clear that each potential forcing may be unambiguously associated with a particular EOF. EOF4 has been included to show that for this eigenvector all potential forcings generally have a contribution at least an order of magnitude smaller than for eigenvectors EOF1 to EOF3.
The eigenvector EOF1, corresponding to the largest eigenvalue, is correlated with NAO, regardless of  whether we are considering transects $58$ and $190$, or all measured bathymetric profiles. Similarly, EOFs 2 and 3 are correlated with MWH and MWL respectively in both cases. Also, the magnitudes of the components corresponding to potential forcings in EOFs 1 to 3 are very close (in fact, they are equal for EOF1 and EOF2 at the level of accuracy considered here) whether we consider the two transects or all the bathymetric measurements together. Hence, the potential forcings are correlated in the same way with transects $58$ and $190$ as they are with the full set of measurements of the bathymetry and so we can use the latter for subsequent analyses. 

The MEOF analysis indicates that the first 3 EOFs are strongly linked to the three forcings considered and together they resolve $98.25$\% of the variance. Note how large the resolved variance is compared to the values of the correlation coefficients. This shows that even though the correlations are small, correlations with any other phenomenon would be significantly smaller and thus we can confidently say that NAO, MWH and MWL are the only relevant potential forcings. Also, since these three EOFs resolve so much of the variance, their associated PCs are used in the MSSA as input channels instead of the original dataset. This simplifies the MSSA considerably as the number of input channels is reduced from 1128 to 3.

 The first 14 MSSA components of the filtered system -- that is, the system consisting of the first 3 PCs as input channels, were analysed.  Each of these components has its corresponding eigenvalue,  spatiotemporal PC (ST-PC) and spatiotemporal EOF (ST-EOF). The criteria presented by \citet{PlautVautard1994} were used to identify coherent patterns within the ST-PCs. The method consists of finding pairs of consecutive eigenvalues  which pass a number of tests; the pairs passing such tests correspond to a quasi-oscillatory pattern. The technique is described in more detail in the Appendix. 
 It was found that all 7 pairs considered passed the test. No more pairs were analysed as these pairs resolve most of the variance.

Once pairs of eigenvalues corresponding to embedded quasi-oscillations are identified, the periods of these quasi-oscillations  can be assessed. The first two pairs correspond to a yearly and a semi-yearly cycle (see Fig.~\ref{fig:MSSA_M121_PCsAtChannel1_Bath_NAO_MSL_MH}; the time span is 120 months, or 10 years), and MSSA confirms that these cycles are correlated with the monthly wave heights as their amplitude is largest at channel~2, which corresponds to MWH. The other pairs have largest amplitude at channel~1, hence they are all correlated with the NAO. The first 4 of these pairs are linked to cycles of period of 3.92 mths, 3.14 mths, 8.62 mths and 2.96 mths. The last pair is associated with two cycles, of 5.4 mths and 2.98 mths. It is noteworthy that no collective interannual patterns seem to exist, even if such patterns have an important influence locally.

\section[Discussion]{Discussion} \label{sec:discussion}

\subsection[sensitivity analyses]{Sensitivity to changes in window length for SSA}
\label{sec:sensitivity}

\subsubsection{Transect 58:}

At this transect a window length of 4 years chosen for the detrending analysis. However, the robustness of the results to small changes in window length was tested by varying the window between 3.3 and 6.6 years (so the window length span includes the sandbar life cycles observed north and south of the pier). It was found that the extrema had approximately the same values for all the window lengths considered, with some small deviations of their location along the transect depending on the window length considered. . 
The main difference observed occurred in the region of small variance, i.e. between $180$ and $220$~m offshore, where the trend signal did not capture any of the extrema when $M=3.3$~years. However, window lengths between $4$ to $6.6$~years extracted trends that have similar characteristics.

The LPOs and SPOs within the two parts of the original signal thus obtained were then identified with SSA at a window length of $9$~years and a window length of $8.12$~years. Again, the dominant patterns were found to be robust to changes in window length, but the window length of $9$~years captured some of the fluctuations better.

\subsubsection{Transect 190:}

Since the window length of $4$~years was best at transect~$58$ detrending was performed initially at this window length. The trends shown in Fig.~\ref{fig:detrending} are at this window length for most of the transect except between $130$ and $170$~m offshore. In this region the $4$~year window did not resolve the travelling patterns adequately, so it was necessary to experiment with slight changes to window length to obtain a coherent picture. The amount of variance resolved by the patterns, however, is the same as that with a window of $4$~yrs, and the characteristics of the travelling patterns are therefore robust (because these slight changes do not introduce more information). It seems that the sensitivity in window length in this region is due to the change in type of behaviour between the region closer to the shore and the patterns travelling seawards between $140$ and  $270$~m (see the patterns developing from year 15 onwards  in Plot b,ii) of Fig.~\ref{fig:detrending}, for instance).

\subsection[Comparison with previous studies]{Comparison with previous Duck studies}
\label{sec:comparison_previous_studies}

Our results compare well with the wavelet analyses presented by
\citet{ReeveEtAl2007}, which focused on temporal scales
of variability at Tr.~$62$. We shall use our results at Tr.~$58$ for the
comparison, since Tr.~$62$ lies only around $100$~m south of Tr.~$58$ and hence 
sufficiently nearby for variations in underlying oscillations to be small.

\citet{ReeveEtAl2007} found that between $100$ and $190$~m offshore most of the variance corresponded to periodicities between $8$ and $12.8$~months. This agrees well with the yearly pattern found at Transect~$58$; however, our trend analysis has highlighted that LPOs are also present in this region, with periodicities varying between $3$ and $12$ years depending on the location. SSA at $260$~m led to three dominant quasi-oscillations of periods 4.5-5.9 years, 1.6-2 years and 3.6-4.8 years. This agrees relatively well with \citet{ReeveEtAl2007} who found
that at $x=260$~m, $25.95\%$ of the variance corresponded to
periodicities between $1.3$ and $1.8$ years, which would be linked to the 1.6-2 yearly patterns identified at Transect $58$.  It is noteworthy
that at $x=410$ m the pattern with a period of $6-8$~years identified
by \cite{ReeveEtAl2007} was not observed.  Therefore, the alongshore homogeneity assumption breaks to some degree at this position. This does not significantly affect the analysis presented here since it is not based on this assumption.

The large variability of temporal scales identified within the dynamics is a consequence of the extremely non-stationary behaviour of beach profiles at Duck. Nonstationarity of the patterns, observed by Reeve et al., is also evident with SSA and had   been noted by \citet{LiEtAl2005} in an earlier wavelet study at Duck. \citet{YiouEtAl2000} have pointed out that  SSA and wavelets can both identify nonstationary behaviour in time series, contrary to more standard time series analysis methods. Also, although wavelets have been the method of choice for intermittent time series analysis, a multi-scale SSA may be produced using a moving window length and the eigenvectors of the resulting lag-correlation matrix are equivalent to data-adaptive wavelets.  
While wavelet analysis may appear conceptually simple, it is not without difficulties. For instance, the Heisenberg uncertainty principle implies that the optimizing wavelet needs to satisfy a trade-off between localisation in time and in frequency \citep{YiouEtAl2000,GrinstedEtAl2004}. It would be interesting to analyse correlations between the bed level behaviour with cross-wavelets between the bed level variations and the suggested potential forcings, at individual locations, in particular because this would permit the identification of the relative phases between the two time series, which is not straightforward with SSA.  This would provide an additional measure of the correlation between the bedlevels and the phenomena time series at given locations.

Sequential beach changes performed by \citet{Birkemeier1984} and \citet{LippmannEtAl1993} can be compared to the results presented
here. Birkemeier's study  covered the period between 1981 and  July 1984, i.e from $t=0$
to $t=3$~yrs, while Lippmann et al.'s study took place between October
1986 and September 1991, i.e. from $t=5.35$ to $10.17$~yrs. In both
studies the shoreline moved on and offshore, always lying between $x=100$m and $x=140$~m. This agrees well with the first spatial region identified with SSA at both window lengths, which implies that these quasi-oscillatory patterns are linked to those of the shoreline.

There is also good agreement between Birkemeier and Lippmann et al.'s studies and this one in relation to the region identified as most dynamic. In this study it is found, for instance, that between $130$~m and $350$~m several quasi-periodic cycles may dominate the dynamics (see Plots (a,ii) and (a,iii) as well as Plots (b,ii) and (b,iii) in Fig.~\ref{fig:detrending}); Birkemeier and Lippmann et al. also found that this was the most dynamic region in terms of number of bars and in terms of bar motion. In their study, the inner bars moved on and offshore between $130$~m and $330$~m, and the outer bars generally between $200$ and $450$~m. Moreover, \citet{LippmannEtAl1993} showed that the inner bar oscillates seasonally around a fixed position of around $200$~m, while \citet[][as cited by \citet{LippmannEtAl1993}]{Birkemeier1984} showed that at transects 62 and 188 the inner bars had significant intra-annual variations and remained generally shorewards of $200$-$210$~m offshore. In this study it is also found that several intra-annual patterns exist shorewards of $200$-$210$~m, as shown in Plots (a,iii) and (b,iii) of Fig.~\ref{fig:detrending}.

\subsection[Potential forcing mechanisms based on temporal patterns]{Underlying quasi-periodic patterns in potential forcing mechanisms based on SSA and possible correlations with bathymetric patterns}
\label{dis_sec:wave}

It is expected that a yearly cycle will be embedded within the wave data and hence may potentially be correlated to the bathymetric yearly cycle. To investigate this, we used the wave data from a wave gauge array, consisting of $15$~pressure gauges 
mounted $0.5$~m off the sea floor, at a
depth of around $8$~meters. The array is located between $835$ to
$955$~m offshore and between $735$ and $990$~m alongshore, i.e. in the
vicinity of transect~$58$. A rose plot of wave directions (not shown)
indicates that the wave crests at this position are almost parallel to
the shore, as all the monthly averaged wave directions are between
$54^o$ and $108^o$ North, with $68.5\%$ of them in the $73^o$-$90^o$
range. Moreover, at this location the mean wave period is around
$9$~s, with a standard deviation of $0.76$~s; thus, there is little
variation in the monthly averaged wave period, $T_p$. Therefore, the
only wave parameter showing significant variation at this location is the wave
height. Since the bathymetric analysis is performed on monthly
interpolations of the data and our interest lies in the long-term (yearly to decadal) evolution of the seabed, we deem  it sufficient to consider the monthly averaged wave heights at this location. This is equivalent to utilising appropriate means in the wave forcings in relation to the timescales of the studied bathymetric change \citep{Cayocca2001}. 

\begin{figure}
\centering   
\epsfig{file=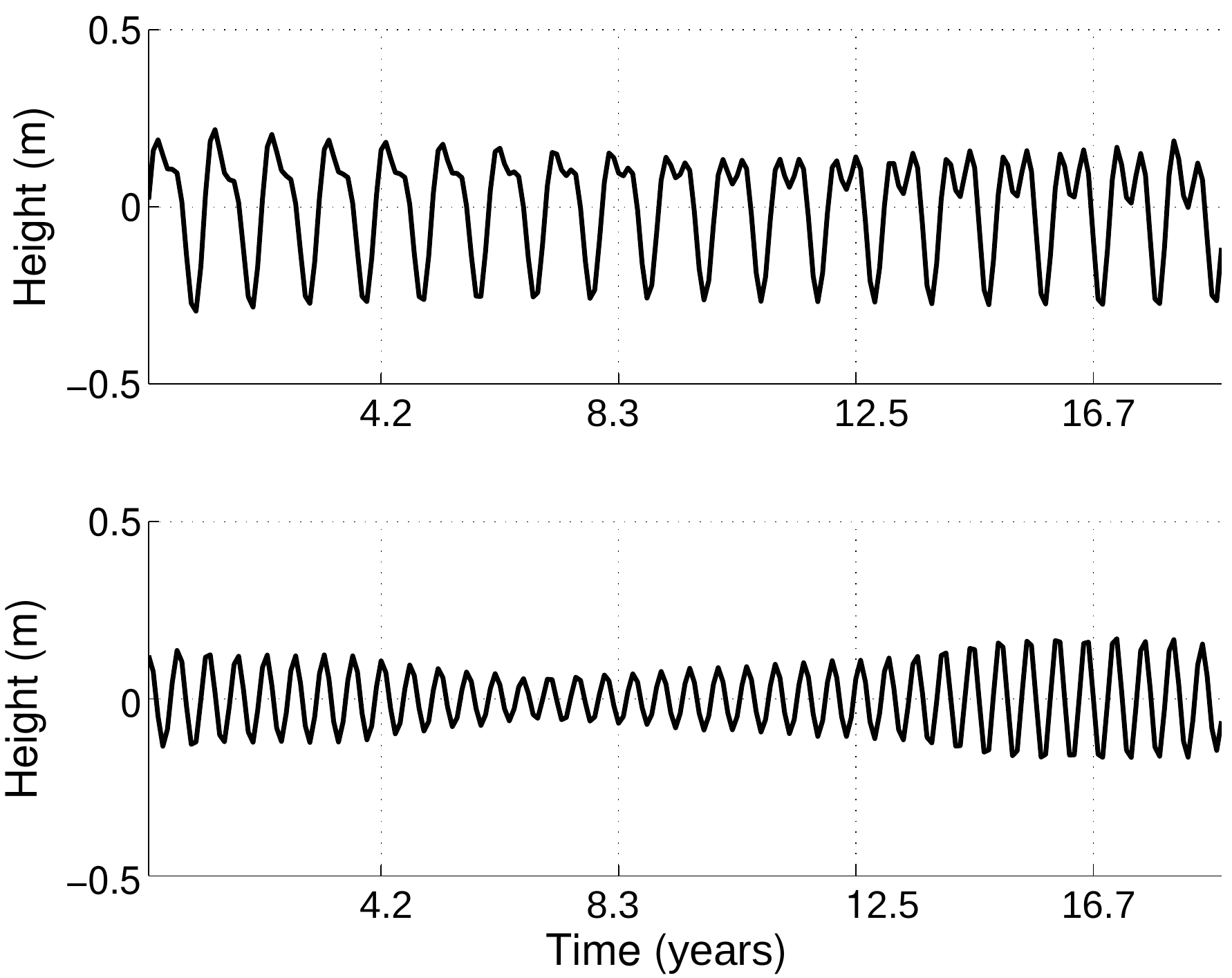,width=8cm}
\caption{SSA reconstruction of wave dynamics, with the first fundamental pair shown at the top and the second fundamental pair shown at the  bottom.}
\label{fig:waves_ssa_reconstruction_1rst_and2nd_pairs}
\end{figure}

SSA of the monthly averaged wave heights clearly indicate the two main underlying quasi-oscillations  are predominantly annual and semi-annual, respectively; these two oscillations are shown in Fig.~\ref{fig:waves_ssa_reconstruction_1rst_and2nd_pairs}. We hypothesise, therefore, that the semi-annual and the annual cycles found in the bathymetric profiles are very likely to be induced by the monthly-averaged wave climate. This is corroborated by the MEOF and the MSSA studies discussed in Sect.~\ref{dis_sec:forcingsSVDandMSSA} and confirms extensive work by previous authors \citep[][and references therein]{LippmannEtAl1993,PlantEtAl1999,LarsonKraus1994,HorrilloCaraballo_Reeve2008}, who have identified the wave climate as one of the dominant forcing mechanisms of nearshore dynamics.

As mentioned in the introduction, storm surges have been shown to have an important influence on coastal erosion. Although the variation of water depth throughout the bathymetry is unknown, a measure of the local water depth, or the mean water level, MWL, is given by tidal gauge measurements at the site, collected by NOAA. An SSA of the MWL highlights two interannual oscillations below 9 years, with periods of 5.7-6.2~yrs and 1.89-2~yrs. Four intra-annual patterns are identified too, with periods of 2.47-2.53~mths, 3.9-4.1~mths, 5.8-6.1~mths and one year. Some of these patterns are similar to those embedded within the bathymetry at some locations.  Note that Kirby and Kirby (2008) identify neap-spring tide cycles and seasonal cycles within tidal mudflats to be linked predominantly to tides and wave conditions, with the bathymetric long-term patterns being strongly linked to NAO. Considering  the interannual MWL patterns we indentified at Duck, MWL may be influencing such long-term patterns in the bathymetry, and not just tidal or seasonal patterns.  However, Kirby and Kirby's study as well as other investigations on the influence of climatic patterns on bedlevel variations would suggest NAO may be influencing bedlevel variations at Duck \citep{RanasingheEtAl2004,KirbyKirby2008,Rozynski2010}.

To check which patterns are embedded within the NAO time series, NAO data was obtained from the coastal research unit of the University of East Anglia (see
http://www.cru.uea.ac.uk/\~timo/datapages/naoi.htm) and analysed with SSA. The annual NAO index between $1850$ and $1999$ was extracted from the dataset and SSA was applied using a window length of $40$ and $50$~years - this allows the identification of interannual as well as some decadal patterns of oscillation underlying the dynamics. The interannual patterns identified are cycles of $7.1$ to $8.8$~years and $2.7$ to $2.9$~years, with a smaller contribution from cycles of periodicity between $3.8$ to $5$~years; another contribution with periodicity between $12$ and $13$~years is also found. These periodicity values are robust to window length changes both in the SSA and the correlogram SDA tests.  
Some of the periodicities within the bed level data coincide with the observed processes but not others. One should not be seeking a physical explanation for all the statistics, as noise produced by, for instance, transverse bars or bar splitting, may cause some variations in the observations \citep{KirbyKirby2008}.  However, conceptual models describing the sediment transport and hydrodynamic processes expected to result in the observed patterns need to be based on links between the patterns so that causalities can be identified. For instance, \citet{Rozynski2010} found that at Lubiatowo there was a weak correlation between NAO and the wave climate. \citet{RanasingheEtAl2004}, on the other hand, found that accretion patterns in the Southern end of Narrabeen Beach (New South Wales, Australia) could be attributed to negative SOI index phases linked to a reduced number of tropical cyclones and East Coast lows in the Australian East Coast which pushed the storms further south leading to a reduction of their effects on the side of pocket beaches sheltered by headlands. In the North Atlantic, the NAO has been shown to produce variations in sea level over decadal timescales \citep{GehrelsLong2007}. At short to medium timescales NAO index phases are likely to be linked to changes in sea level pressure lows and highs and have an effect on the direction and intensity of the storms, in analogy with the observed effects of the SOI in the South Pacific Ocean. Most research so far has identified links between climatic indexes and erosion and accretion patterns, but it is more difficult to assess their effects on the dynamics of NOM systems. Also, at Duck the pier may be introducing patterns similar to those observed on sheltered beaches depending on the direction of the incoming waves. Additional research looking at NAO index phases would be necessary to propose any conceptual model on how NAO may be influencing the dynamics at Duck.   This is beyond the scope of the current investigation which only concerns quantification of cyclic patterns and is left for future studies.
In order to analyse further whether the potential forcings discussed in this section are correlated with seabed level measurements throughout the bathymetry, and which collective patterns of behaviour they may be correlated with, we performed spatio-temporal correlation analyses using MEOF and MSSA and these will be discussed next.

\subsection[Discussion on spatial and spatiotemporal correlations]{Spatial and spatiotemporal correlations between potential forcings and bathymetry and possible causalities} \label{dis_sec:forcingsSVDandMSSA}

As shown in Sect.~\ref{res_sec:meof_mssa}, the MEOF has permitted the isolation of the effects of the different potential forcings on the bathymetry because each potential forcing is most strongly correlated with one of the first 3 EOFs: the NAO with EOF1, the monthly wave heights (MWH) with EOF2 and the monthly mean water level (MWL) with EOF3. 

\begin{figure}[t]
\centering
\subfigure[\label{fig:contours_Correlations_NAO} NAO Correlations]{\epsfig{file=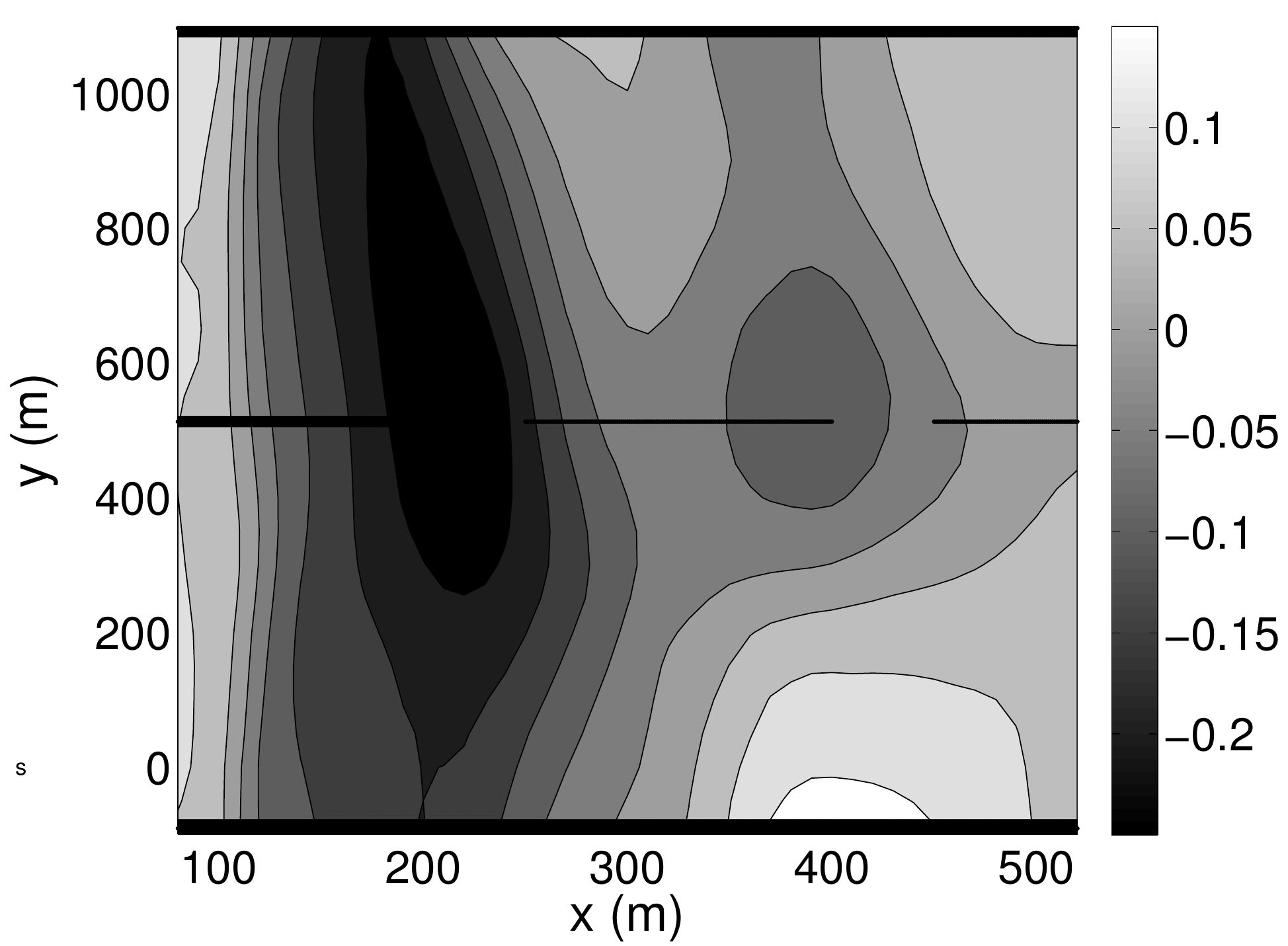,width=4.4cm}}
\vspace{0cm}
\subfigure[\label{fig:contours_Correlations_MWH}  MWH Correlations]{\epsfig{file=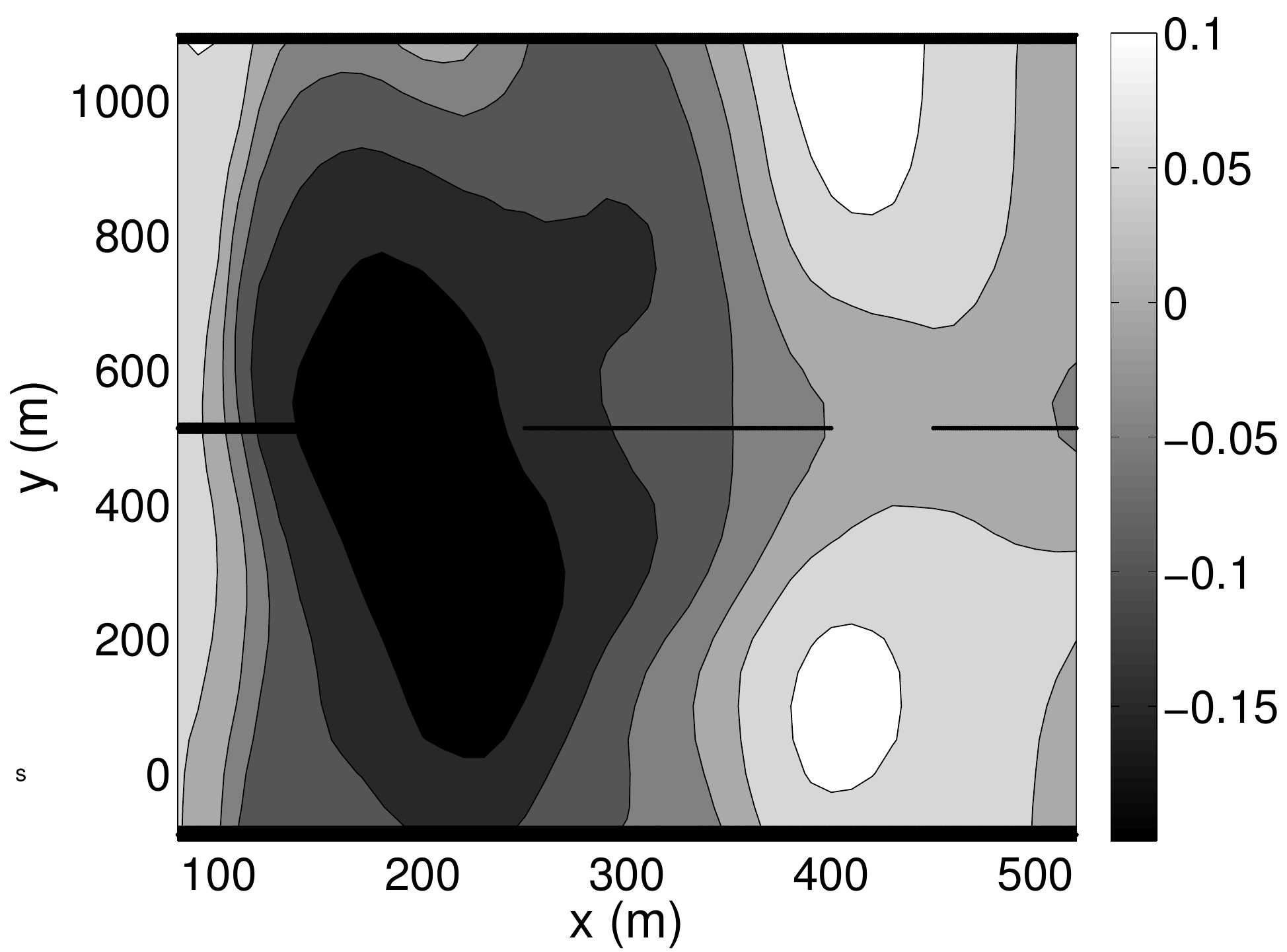,width=4.4cm}}
\subfigure[\label{fig:contours_Correlations_MWL}  MWL Correlations]{\epsfig{file=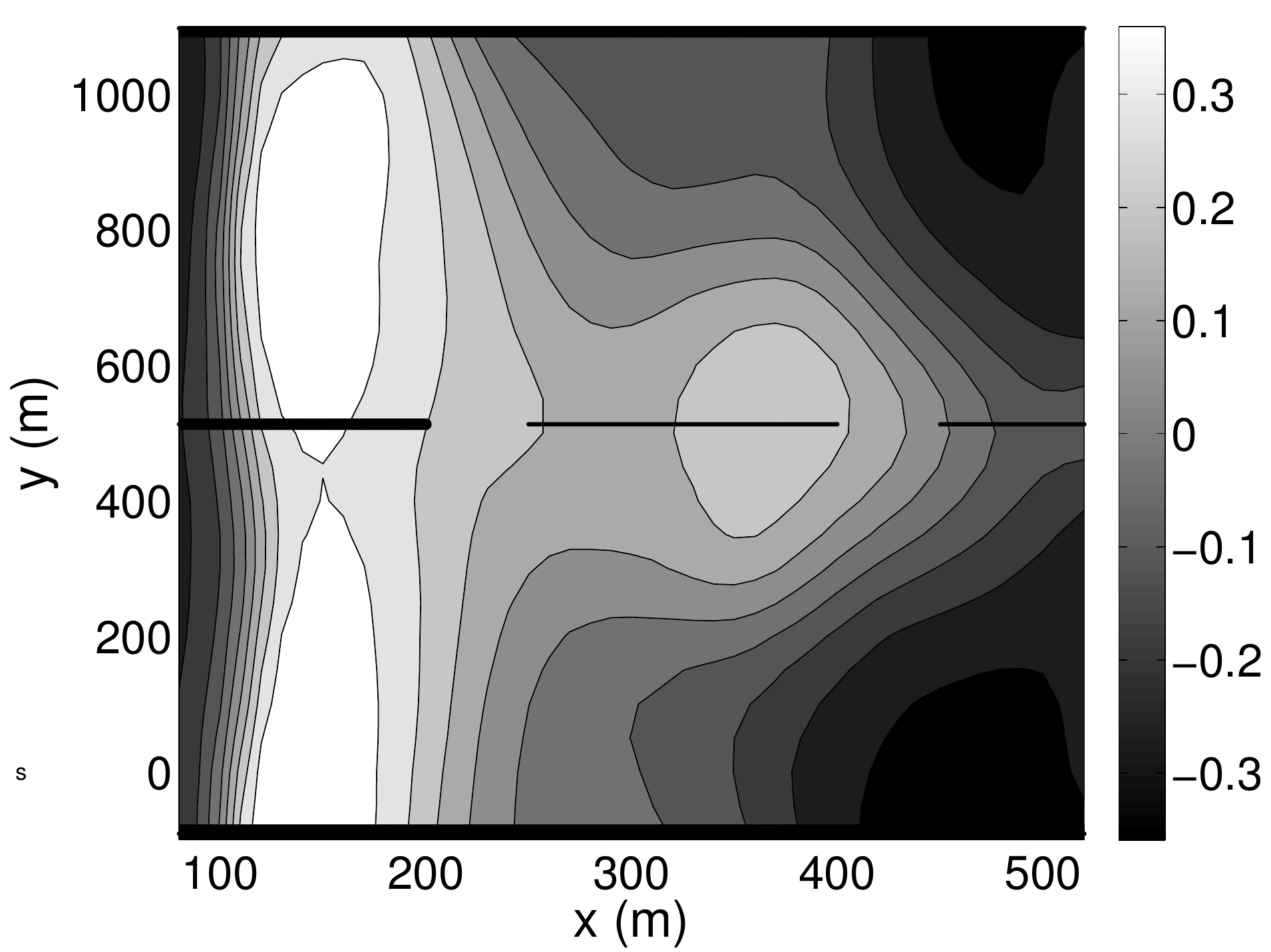,width=4.4cm}}\\
\vspace{0cm}
\caption{\label{fig:Correlation_Contours} Correlation Contours
between the seabed and the NAO, the MWH and the MWL, respectively. Tr. 58 (top, Northern side) and Tr. 190 (bottom, Southern side), are highlighted (solid lines). The pier is located at $y=513$~m (dashed line). }
\end{figure}

Figs.~\ref{fig:Correlation_Contours} permit the identification of regions where each of the potential forcings is more strongly correlated with seabed evolution, simply by locating the regions where the correlations are highly positive or highly negative. Note that the region where the analysis was performed extends from $80$ to $520$~m offshore, rather than between $100$ and $490$~m, as was the case in the previous sections. As explained before, this was done to capture the behaviour around the shoreline and the behaviour offshore more fully.

Since the first EOF captures the largest proportion of the variance and it is linked to the NAO, it appears that the NAO is more strongly correlated with the seabed than are the waves at the time and space scales under consideration. Plot~\ref{fig:contours_Correlations_NAO} shows the distribution of correlations between the seabed and NAO, and highlights the complex interactions between them; the regions of positive and negative correlation are clearly visible in the plot. It shows the correlations are either very weak (with $|r|<0.1$) or weak (with $0.1<\left\vert  r \right\vert \leqslant  0.3$). Weak correlations occur 1) North of the pier between 600~m and 1100~m alongshore and 80 to 100~m cross-shore; 2) South of the pier between -100~m and 100~m alongshore and 360~m and 490~m cross-shore and; 3) all along the shore between approximately 150~m and 240~m cross-shore. However, even though the correlation is low, it is found that the correlation coefficient contours give valuable information and the locations of extrema (either positive or negative) are more important that the magnitude of the extrema. An analysis of extrema close to the pier shows, for instance, that there are stronger correlations between NAO and the seabed between $180$ and $290$ and between $350$ and $450$~m offshore than at other cross-shore locations. Moving northward from the pier these regions, while remaining distinct, move closer to the shore. Thus, at transect~$58$, the largest negative correlations occur now  between $150$ and $200$ and between $320$ and $380$~m offshore (approximately), with two regions of positive correlations appearing between $250$ and $280$~m  and seawards of $450$~m. Southwards of the pier, the region of negative correlation is around $150$ and $230$ and a region of positive correlations appears between $350$ and $450$~m. Note that although the magnitude of the correlations is low the variance resolved by the EOF associated with the NAO is large (see Table~3) (remember that the correlation coefficient is linked to the spatial EOFs of the MEOF covariance matrix).  The variance is a relative measure indicating that along this direction the correlations with other phenomena will be much weaker. 

The regions where NAO and the bathymetry are more likely to be correlated (or anti-correlated) relative to other seabed locations - even if in absolute terms the correlations are everywhere weak - and the correlations between them deduced from the SSA at transects $58$ and
$190$ seem to be consistent with one another. For instance, Table~2 shows that, for transect $58$, regions seawards of $150$~m have one underlying cycle at a periodicity found within the NAO signal: as we know, the NAO has 3 interannual underlying cycles with periods of around, 2, 5 and 6-8 years, respectively, and if for instance we choose the $250$m offshore location, we see that the period of 5yrs is contained within the bathymetric pattern with period 5.4$\pm$0.9; at $480$~m and $490$~m  two of the dominant bathymetric cycles
identified have periods that are close to two dominant patterns
within NAO, namely the 2 and the 6-8 yearly cycles. For transect $190$, patterns of the same periodicity as those of the NAO were found both from $120$ to $200$~m and beyond $450$~m offshore. A possible physical mechanism for the NAO to influence the bathymetry is indirectly through the thin layer of water in the nearshore, producing cyclic patterns of behaviour in the bathymetry with periodicities close to its own underlying cyclic patterns.

The MWH is the dominant potential forcing in EOF2 and the distribution of correlations between MWH and the seabed is shown in plot~\ref{fig:contours_Correlations_MWH}. Again, the regions around positive and negative correlation extrema indicate regions where the MWH is more likely to be correlated with the seabed oscillations. There is agreement between the regions identified here and those where the waves may have an important effect: for instance, in the swash zone, i.e. between $80$ and $90$~m offshore, where there are positive correlations. Looking at the SSA results in more detail, we know from the SSA that the wave height has a yearly pattern and that such a pattern is also present shorewards of $310$~m at transect $58$, and all along the transect at transect $190$. As discussed above, the correlation contour  plot~\ref{fig:contours_Correlations_MWH} seems to indicate, indeed, that  the correlation between the seabed and the waves near the pier and  close to the shoreline is stronger than at other bathymetric locations, as evidenced by the region of negative correlations occurring between $130$ and $250$m offshore and between $0$ and $800$~m alongshore,  approximately.  As we approach transect
$190$, there appears to be a  correlation minimum between $200$ and $260$ and  a correlation maximum between $380$ and $500$ approximately, even though the correlations are overall weak. On the other hand, at transect $58$ there seem to be more variations in the correlation values, which may  indicate there is more variability in the interaction between the waves and the seafloor along  this transect than along transect $190$.

Finally, plot~\ref{fig:contours_Correlations_MWL} shows the distribution of correlations between the seabed and the MWL.  Plot~\ref{fig:contours_Correlations_MWL} show there are regions of moderate correlation and the distribution of correlation coefficients is symmetric around the pier. The plots suggest the greatest correlation with the mean water level could be within
$120$-$200$~m and $380$-$500$~m along transects 58 and 190. This is relatively consistent with the locations where a
$1.4$-$2.2$~yearly cycle was found (see Table~2) and, thus, with the
wavelet findings of \citet{ReeveEtAl2008}. This work not only confirms the presence of those  bathymetric patterns but also suggests the physical mechanism influencing them is one which, at least partially, contributes to changes in water level variations. These could be any of the storm surge components.

The rest of this section focuses on the MSSA results. As presented in Sect.~\ref{res_sec:meof_mssa}, no interannual coherent, spatio-temporal patterns are found when all the bathymetric measurements and the three potential forcings are analysed together, even when locally these patterns are clearly identified. This is an important difference with the SSA/MEOF analyses, where the local bathymetric patterns could be attributed to any of the three forcings considered, depending on the characteristics of the patterns and the region under consideration. However, the absence of coherent interannual, spatiotemporal patterns possibly means that the mechanism forcing such behaviour acts locally and not globally at these temporal scales.

For studies including monthly responses, the results in Sect.~\ref{res_sec:meof_mssa} show that the two most important collective patterns, the yearly and the semi-yearly cycle, are both correlated with the monthly wave heights. This agrees well with the SSA results summarized in Table~2, showing the ubiquituous presence of the annual cycle all along Transects~$58$ and~$190$. Finally, except for the semi-annual pattern all the  intraannual collective patterns identified are correlated with the NAO.

\section{Summary and Conclusions}
\label{sec:conclusions}

In summary, local and collective bathymetric quasi-periodic patterns of oscillation were identified from monthly profile surveys spanning 26 years, from July 1981 to January 2006, at the USACE field research facility in Duck, North Carolina. The local analyses were based mostly on the outmost transects of the surveyed domain, but the collective patterns were computed using all surveyed bathymetric profiles, including those in the vicinity of the pier.  Correlations with three potential forcings,  namely the monthly wave heights (MWH), monthly mean water levels (MWL) - which include the storm surges as well as the tide - and the large scale atmospheric index the North Atlantic Oscillation (NAO), were discussed.  

The local seabed patterns and their spatial correlations with NAO, MWH and MWL were first analysed using SSA, which showed the highly nonstationary behaviour of the bedlevels along transects 58 and 190. It was shown 1) that a quasi-yearly cycle was embedded at most bathymetric locations and this is likely to be correlated with the quasi-yearly pattern  within the monthly averaged wave climate; 2) the local behaviour is highly nonstationary as observed by previous studies;  3) patterns with periodicities in the range between $1.4$ and $2.2$ were found at some seabed locations and also within the MWL, providing an explanation for the potential origin of such patterns and 4) several interannual patterns were identified, some of which are likely to be correlated with the NAO.  Some statistics may be attributed to noise. As in other studies on Duck,  differences in the behaviour North and South of the pier were observed.
 
Linear correlation coefficient contours were computed throughout all bathymetric surveys. The regions at the North and South boundaries of the domain where the magnitude of the correlation coefficient was largest (relative to other locations along the transect) seem to agree broadly with those identified with SSA, even though the correlation coefficient gives a measure of linear correlation while SSA is a nonlinear measure; in fact linear correlation analysis cannot characterise the nonlinear quasi-oscillations. It was found that linear correlations with the NAO and the MWH were weak to very weak. For the MWL, the correlation contours were symmetric North and South of the pier and there was a region between $100$ and $200$ metres offshore of moderate correlation between bedlevels and MWL for all the surveys.

The last component of the analysis consisted of the identification of spatio-temporal oscillations and their nonlinear correlation with NAO, MWH and MWL.  This indicated that there are no patterns that are coherent throughout the bathymetry at yearly timescales, but there are a few at monthly scales which are correlated with the NAO. This is consistent with the highly localised nature of the interannual quasi-oscillatory patterns that was observed with SSA. This study is a preliminary step towards, for instance, an understanding of the potential effects of increased storminess (due to climate change) on bathymetric evolution, or attempts to model nearshore long-term morphodynamics.

\section{Acknowledgements}
This work was supported by an RCUK Academic Fellowship from the EPSRC (Grant number EP/C508750/1) and a Research and Innovation Fellowship award from the University of Plymouth, UK. Thanks to Tom Lippmann and David Huntley for useful discussions. The editors and two anonymous referees are thanked for valuable comments.

\bibliographystyle{elsarticle-harv}

\newpage

\appendix

\section{ Description of methods }
\subsection[SSA and SDA]{SSA and SDA }
\label{method_subsec:ssa_sda}

The fundamental theory of SSA has been discussed by \citet{BroomheadKing1986}, \citet{VautardGhil1989} and   \citet{GhilEtAl2002} in relation to discrete time series analysis. The following summarises the SSA methodology. Suppose the bed level variable $z(t)$, sampled at times $t=i \tau_s$, $i=1,2, \ldots,N$ ($N$ being the length of the time series and $\tau_s$ the sampling interval), at a given cross-shore and long-shore location $(x,y)$,  characterises the seafloor dynamical system.  First, the noise part of the bed level time series may be identified by computing the statistical dimension, $S$, which is found by constructing a secondary, $M-$dimensional sequence, $Z_n=\left( z_n, z_{n+1}, \ldots, z_{n+M-1} \right)$, for $n=1,2, \ldots, N-M+1$, where $M<N$ is the window length and $z_n=z(n \tau_s)$. Time scales of the dynamics that can be reconstructed from this time series are between $\tau_s$ and $\tau_w=M \tau_s$, where $\tau_w$ is the window time span. Then, the eigenvalue problem
\begin{equation}
 {\bm{C}}{\bm{e}}^{(k)}=\lambda_k {\bm{e}}^{(k)},
\end{equation}
must be solved, where ${\bm{C}}$ is the covariance matrix of the
sequence $Z_n$ and $\lambda_k$ are the eigenvalues associated with the eigenvector 
${\bm{e}}^{(k)}$. The square
roots of $\lambda_k$ are called the singular values, with their set
being called the {\textit{singular spectrum}}. The largest singular
values relate to the directions that resolve most of the variance, with the rest being a representation of the noise component and appearing as an approximately flat floor in a plot of singular value {\it{vs.}} singular value rank (with the singular values ordered from largest to smallest). The dimension $S$, then, will be given by the number of singular values above this floor.

Now, each singular value $\lambda_k$ has an associated principal component (PC), defined as
\begin{equation}
A_k(i \tau_s)=\sum^M_{j=1} z(\{i +j-1\}\tau_s)e^{(k)}_j.
\end{equation}
Once the PCs have been identified, the part of the original time series related to that PC, or to a combination of $K$ PCs, may be obtained by computing
\begin{equation}
RC(i \tau_s)=\frac{1}{M_i} \sum_{k \in K} \sum_{j=L_i}^{U_i} A_k(\{i -j+1\}\tau_s){e}^{(k)}_j,
\end{equation}
where RC, the {\textsl{reconstructed component}}, is the part of the original signal related to the PC combination. The values of the renormalisation factor, $M_i$, and the lower and upper bound, $L_i$ and $U_i$, respectively, depend on the time at which the reconstruction is being evaluated.

The frequencies that can be resolved with SSA depend on the choice of window length, $M$, which is a free parameter of the method. However, the patterns extracted should be robust to small changes in window length. Here the depth time series were divided into a trend and a detrended signal. The trend generally contains a part found by linear regression of the data, as well as the RC corresponding to the first few eigenvectors of the SSA at a given window length $M$. These first few eigenvectors generally contain  all oscillations within the original time series that have periods above $M$, which in this case was taken as four years.  From the trend thus computed, long-period oscillations (LPOs) were extracted, while from the detrended signal short-period oscillations (SPOs) may be characterised. Cycles of three to five years may fall in either signal because of the window length selected for the detrending analysis (this will be discussed in more detail in subsequent  sections).

In natural systems, there is the additional problem of red noise (which is noise composed of many random frequencies but, in contrast  to white noise where all frequencies have equal intensity, the intensity of the lower frequencies is stronger)  that can be removed from the data by using either Monte Carlo or chi-squared tests. In this work a chi-squared test was applied on $100$ red-noise surrogates; these surrogates were projected onto the SSA eigenvector basis, and the distribution of projections was approximated as chi-squared with $3M/N$ degrees of freedom. Then those projections that fell outside a percentile range of $2.5\%$ to $97.5\%$  were identified as being part of the signal, and could thus be dissociated from the red noise in the time series.

Spectral density analysis (SDA) is a well established technique to analyse oscillatory patterns in time series. The spectral density was computed  using Welch's \citep{Welch1967} periodogram method and the Blackman-Tukey correlogram approach \citep{BlackmanTukey1958}. The former consists of dividing the input signal into short segments and computing a modified periodogram for each segment, thus leading to a set of periodograms which are finally averaged to obtain an estimate of the spectral density function. In the latter a windowed Fast Fourier Transform of the autocorrelation function is equated to the spectral density.

It is well known that the spectral density is a measure of the energy contained in the system. By analysing the spectral density it is possible to characterise how the total energy is distributed and, in particular, to find the frequencies that contribute the most to it.

\subsection{MEOF/Correlation analysis}
\label{subsec:meof_description}

Multivariate empirical orthogonal functions can be used to study the
relationship between different physical variables that are sampled at
the same set of times. 
It can
be thought of as a prefiltering step for MSSA/MEEOF analysis: instead
of performing the MSSA directly on the data, one first finds the MEOFs
and uses time-lagged copies of the corresponding PCs in the MSSA, in
order to reduce the dimension of the covariance matrix. 

We construct a matrix ${\bm B}$  whose elements
\[B_{ij}=x^{(j)}_i\]
are the values of the $j$th variable sampled at time $i\tau_s$, where $i=1,2,...,N$ and $j=1,2,...,L$, where each vector ${\bm x}^{(j)}$ contains the full time series for a particular variable. The full set of variables might include samples of the same physical variable at different locations and/or different physical variables at the same location(s).

We next subtract from each element of ${\bm x}^{(j)}$ the mean value of all its elements, to form a matrix ${\bm F}$ consisting of columns ${\bm y}^{(j)}$ with zero mean:
\[ F_{ij} =  y^{(j)}_i = x^{(j)}_i - \frac{1}{N}\sum_{i=1}^N x^{(j)}_i.\]

Since the columns of ${\bm F}$ represent different physical variables, we
renormalise them so that the results of the analysis are not
influenced by the units in which we choose to measure them
\citep{MoteEtAl2000}. Our data consists of measurements of the
bathymetry at 1125 separate locations over Duck beach
(covering the area between $y=-91$ to $1097$~m along the shore and
between $x=80$ to $520$~m offshore), together with
measurements of the monthly mean water levels (MWL) and monthly wave heights (MWH) at a
wave gauge located near Duck beach, and a measurement of the North Atlantic Oscillation (NAO)
spatially averaged over the whole surveyed area, the NAO being the
pressure difference between Iceland and the Azores. Since all of these forcings were
simultaneously available for only the last 19 years of bathymetric surveys, we reduced the
analysis to this duration. We considered two
possible choices for the renormalisation factor: the range of each
variable or its standard deviation. We decided to use the range
because we were interested in understanding the effect of each
possible forcing independently, and using the range separated these
out in the calculated MEOFs to a greater degree than using the
standard deviation. Finally, since we have 1125 pieces of data for the
bathymetry at each time $i\tau_s$ and only one piece for each possible
forcing, we divide the bathymetry measurements by 1125 so that the
effect of the potential forcings is comparable with that of the spatial
variations in the bathymetry. From now on we consider ${\bm F}$ to have been renormalised
in this manner.

Next, a singular value decomposition (SVD) of ${\bm F}$ permits the identification of bathymetric 
locations where the bathymetry may be strongly correlated with each of the potential forcings. 
Such decomposition is of the form
\[ {\bm F} =  {\bm U} {\bm S} {\bm V}^T, \]
where the columns of ${\bm V}$ are the EOFs of the SVD, or the spatial
eigenvectors, the columns of ${\bm U}$ are the PCs, or the
temporal eigenvectors and ${\bm S}$ is the diagonal matrix of eigenvalues; the ratio of each eigenvalue divided by the sum of all the eigenvalues  gives the fraction of the total variance explained by each eigenvector. The EOFs represent the stationary patterns in
the bathymetry and potential forcings that explain most of the
observed variability, with the greatest variance being resolved by the
EOF corresponding to the largest eigenvalue and so on. The PCs provide the variation over time of the sign and amplitude of the associated EOF.

Each spatial eigenvector giving a direction in which each potential forcing is dominant gives an excellent proxy of the relative correlation of that phenomenon with that particular spatial mode of the bathymetry. In fact, computing the correlations directly from the data and comparing the correlation distributions to the spatial EOF contours would show that correlation contours and EOF contours are equivalent. This is because of the dominance of the phenomena along well defined EOF directions.  Since linear correlations are conceptually more straightforward than EOF distributions, we present and discuss the correlation plots instead of the EOF distributions. However, we use MEOF as a filter for the MSSA.

\subsection{ MEEOF/MSSA}
\label{subsec:meeof_description}

Multivariate Extended Empirical Orthogonal Functions (MEEOF) is the
name given to Multichannel Singular Spectrum Analysis (MSSA) when
several different physical variables are included in the analysis.
Such a technique has been useful, in particular, in the analysis of the behaviour of climatic phenomena, with the aim of extracting modulated oscillations from the coloured noise commonly present in natural systems \citep{VautardGhil1989,MoteEtAl2000}. Its use has focused on accurate identification and characterisation of the system's temporal and spectral properties. MSSA is fully described in \citet{AllenRobertson1996} and references therein. Here a short description is presented for the reader interested in the methodology and its capabilities.

Suppose we have the dataset, $\left\{ y^{(j)}_i;\,i=1,\ldots,N, \, j=1,\ldots,L \right\}$, discussed in the previous section.
Several techniques to produce a matrix of time-lagged vectors may be
proposed. Here the method used by \citet{AllenRobertson1996} is adopted. It consists of creating the matrices $\tilde{\bm{Y}}^{(j)}$, formed by setting the channel to $j$, $1 \leq j \leq L$, sliding a vector of length $M$ down from $i=1$ to $i=N' \equiv N-M+1$ and ordering the resulting vectors from left to right as the columns of $\tilde{\bm{Y}}^{(j)}$. Thus each $\tilde{\bm{Y}}^{(j)}$ has size $M \times N'$. This leads to a \textit{trajectory matrix},
\begin{equation}\label{eq:trajectory_matrix}
\tilde{\bm{Y}}=\left( \begin{array}{c} {\tilde{\bm{Y}}}^{(1)}\\ \vdots \\ {\tilde{\bm{Y}}}^{(L)} \end{array} \right)
\end{equation}
of size $ML \times N'$.

 A singular value decomposition of $\tilde{\bm{Y}}$ may then be performed, such that
\begin{equation} \label{eq:svd_X}
\tilde{\bm{Y}}= \bm{P}_{\tilde{\bm{Y}}} \bm{\Lambda}_{\tilde{\bm{Y}}}^{1/2} \bm{E}_{\tilde{\bm{Y}}}^{T},
\end{equation}
where $\bm{P}_{\tilde{\bm{Y}}}$ consists of the temporal principal components (T-PCs). As equation \ref{eq:svd_X} implies, these are the left-singular vectors of ${\tilde{\bm{Y}}}$. The matrix $\bm{E}_{\tilde{\bm{Y}}}$ consists of the spatio-temporal empirical orthogonal functions (ST-EOFs) corresponding to the right-singular vectors of ${\tilde{\bm{Y}}}$. Finally, $\bm{\Lambda}_{\tilde{\bm{Y}}}$ is the diagonal matrix of variances associated with these orthogonal bases \citep{Robertson1996}.

The bases $\bm{P}_{\tilde{\bm{Y}}}$ and $\bm{E}_{\tilde{\bm{Y}}}$ are the eigenvectors of the covariance matrices
\begin{equation} \label{eq:covariance_matrices}
\bm{C}_{\tilde{\bm{Y}}}^P=\frac{1}{ML}{\tilde{\bm{Y}}} {\tilde{\bm{Y}}}^T \mbox{ and } \bm{C}_{\tilde{\bm{Y}}}^E=\frac{1}{N'}{\tilde{\bm{Y}}}^T {\tilde{\bm{Y}}},
\end{equation}
respectively. However, it is only for the smallest of $\bm{C}_{\tilde{\bm{Y}}}^P$ and $\bm{C}_{\tilde{\bm{Y}}}^E$ that either $\bm{P}_{\tilde{\bm{Y}}}$ or $\bm{E}_{\tilde{\bm{Y}}}$ is full-rank \citep{AllenRobertson1996,Robertson1996}. Therefore, the choice of $M$ determines which of the eigenbases is to be analysed. However, the method of construction of $\tilde{\bm{Y}}$ implies that the column vectors of $\tilde{\bm{Y}}^T$ are time-lagged copies at a given channel, so it is convenient to choose $N'$ small enough to satisfy $N'<ML$, so that the full rank matrix is $\bm{C}_{\tilde{\bm{Y}}}^E$. 

As mentioned in Sect.~\ref{subsec:meof_description}, rather than performing the MSSA over the data and having to resolve a very large covariance problem, it is better to perform a data reduction procedure by prefiltering the original data with the MEOF and apply the MSSA to the resulting EOFs. In this way, the most important information is compressed into a small number of variables \citep{PlautVautard1994} and the analysis of the coherent patterns is simplified significantly. In this case the first three eigenvalues of the MEOF resolve most of the variance and also are the ones in the directions of the three potential forcings considered, so the MSSA can be applied to these three eigenfunctions rather than to the data.

The first  14 MSSA components of the filtered system -- that is, the system consisting of the first 3 MEOF  PCs as input channels, were analysed.  Each of these components has its corresponding eigenvalue,  spatiotemporal PC (ST-PC) and spatiotemporal EOF (ST-EOF). A window length of $M=120$, equivalent to $N'=9$~years, was chosen. The criteria presented by \citet{PlautVautard1994} were used to identify coherent patterns within the ST-PCs. The method consists of finding pairs of consecutive eigenvalues $(k,k+1)$ for which:
\begin{itemize}
\item the two eigenvalues are nearly equal, that is they are both within the margins of error identified by North's rule of thumb \citep{NorthEtAl1982},
\item the two corresponding time sequences described by the ST-PCs are nearly periodic, with the same period and in quadrature,
\item the associated ST-EOFs are in quadrature and
\item the lag correlation between ST-EOFs of order $k$ and $k+1$ itself shows oscillatory behaviour \citep{GhilMo1991}.
\end{itemize}
Pairs that pass all these tests are associated with quasi-oscillations with periods equal to the period of the ST-PCs - which is also the period of the ST-EOFs - in the embedding space, here spanning 121 months.  

The first test was to inspect pairs of consecutive eigenvalues $\lambda(k),\lambda(k+1)$ to see if they satisfied North's rule of thumb. It was found that eigenvalue $\lambda(k+1)$ lies within the interval $[\lambda(k)(1-\sqrt{2/N});\lambda(k)(1+\sqrt{2/N})]$, with $N=228$~months being the sampling duration, for all pairs considered, so the 7 of them satisfy the first criterion. Then plots of the ST-PCs for the 14 MSSA components at channel~1 showed that all pairs considered are quasi-periodic with same period and in phase quadrature so they also satisfy the second criterion. This led to a check of the quadrature criterion for the ST-EOFs associated with each ST-PC pair. Since the ST-EOFs are the same for all the PCA channels in the prefiltering, only the results for channel~1 are necessary. The ST-EOF plots are shown in Fig.~\ref{fig:MSSA_M121_PCsAtChannel1_Bath_NAO_MSL_MH} for the first 6 pairs, with the solid line corresponding to the eigenvector with smaller index. All the pairs (including pair 13-14, not shown) pass the test. The final test consists of analysing the lagged correlation plots for the pairs which have satisfied all previous tests. The plots, shown in Fig.~\ref{fig:MSSA_M121_PCPairs_LagCorr_Bath_NAO_MSL_MH}, should depict oscillatory functions. Again, all pairs pass this test.

\newpage

\noindent {\large{\bf{Figure Captions:}}}

\noindent Caption 1: Data contours for transects 58 (top, North of the pier) and 190 (bottom, South of the pier) with: beach elevation (in metres) displayed in Figs.~(a,i) and (b,i); trend signal elevation (in metres) displayed in Figs.~(a,ii) and (b,ii) and; detrended signal elevation (in meters) displayed in Figs. (a,iii) and (b,iii). The upper panels in Figs.~(a,ii), (a,iii), (b,ii) and (b,iii) show
  the percentage of variance resolved by the corresponding signal.

\noindent Caption 2: Quasi-oscillations identified within the trend (left) and the detrended (right) signals together with the resolved variance, at window lengths of $9$~years, for transects $58$ (top) and $190$ (bottom).

\noindent Caption 3: Spectral density plot for the reconstructed component of the SPOs at $x=160$~m for Transect~$58$. The thin lines indicate the $95\%$ confidence interval for the spectral estimates.

\noindent Caption 4: SSA reconstruction of wave dynamics, with the first fundamental pair shown at the top and the second fundamental pair shown at the  bottom.

\noindent Caption 5: Correlation Contours
between the seabed and the NAO, the MWH and the MWL, respectively. Tr. 58 (top, Northern side) and Tr. 190 (bottom, Southern side), are highlighted (solid lines). The pier is located at $y=513$~m (dashed line).  

\noindent Caption 6: First six pairs of ST-EOFs of the MSSA components linked to potentially quasi-oscillatory pairs at all channels, with their associated resolved variance. The abscissa represents time (in months) and spans the window length $M=121$~months.

\noindent Caption 7: ST-PCs for consecutive MSSA components at channel~1; the ST-PCs span the window length $N'=108$~months.

\noindent Caption 8: Lagged correlations for all ST-PC pairs.

\newpage
\noindent {\large{\bf{Figures:}}}

\begin{figure}[ht]
\subfigure[Pair 1-2]{\label{fig:MSSA_M121_STEOFs_Pair_1and2_Bath_NAO_MSL_MH}
\includegraphics[width=5.5cm]{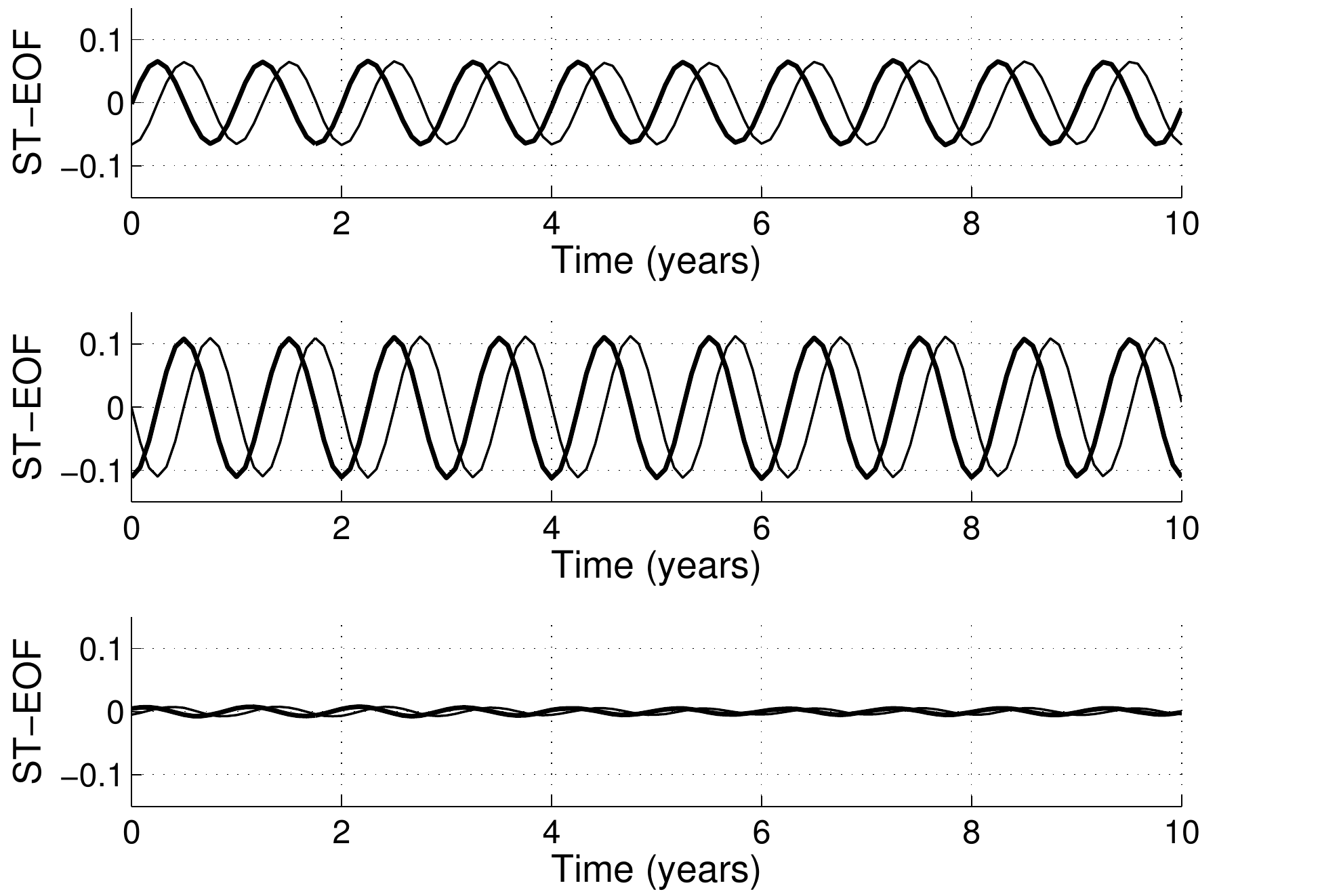}}
\subfigure[Pair 3-4]{\label{fig:MSSA_M121_STEOFs_Pair_3and4_Bath_NAO_MSL_MH}
\includegraphics[width=5.5cm]{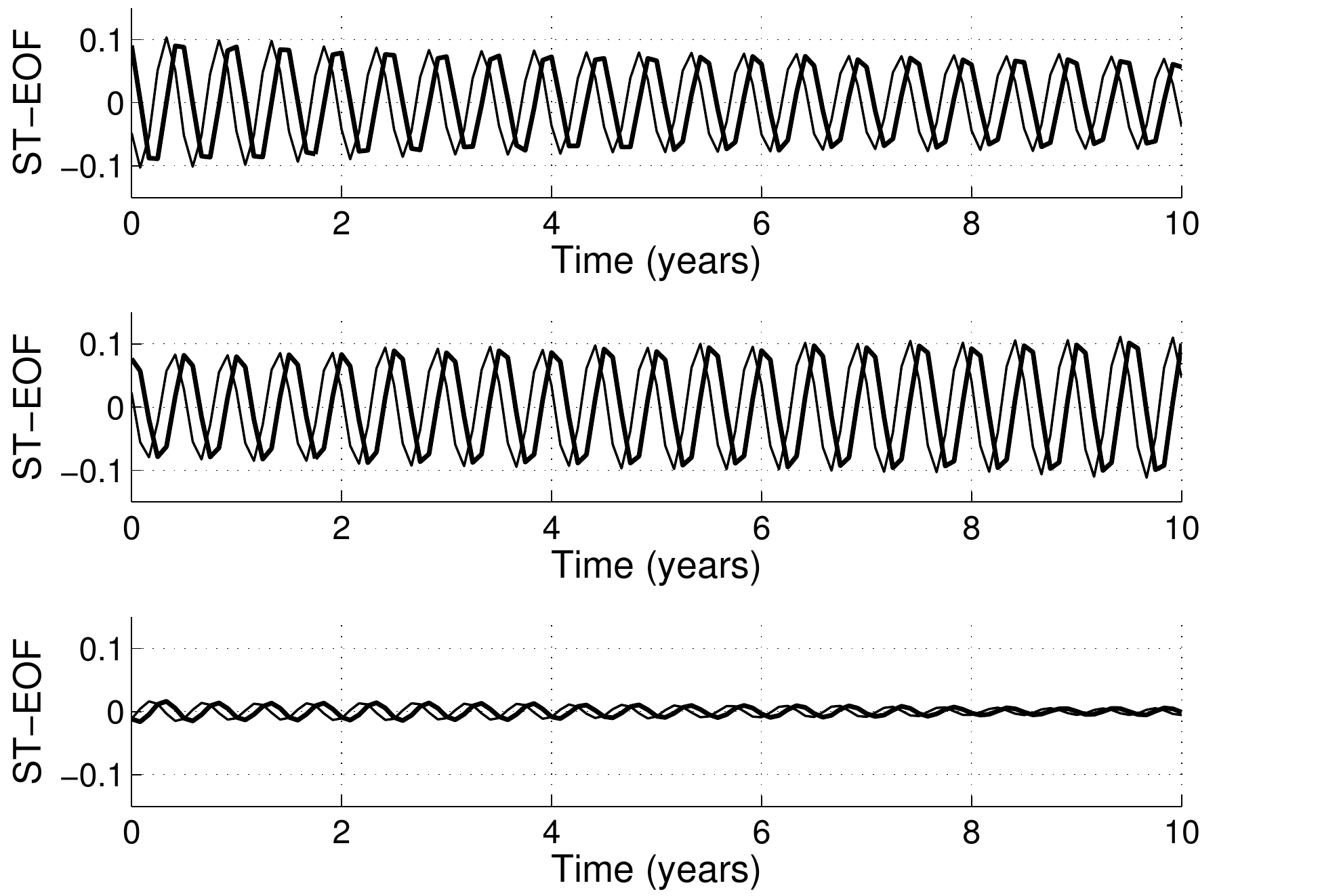}} \\
\subfigure[Pair 5-6]{\label{fig:MSSA_M121_STEOFs_Pair_5and6_Bath_NAO_MSL_MH}
\includegraphics[width=5.5cm]{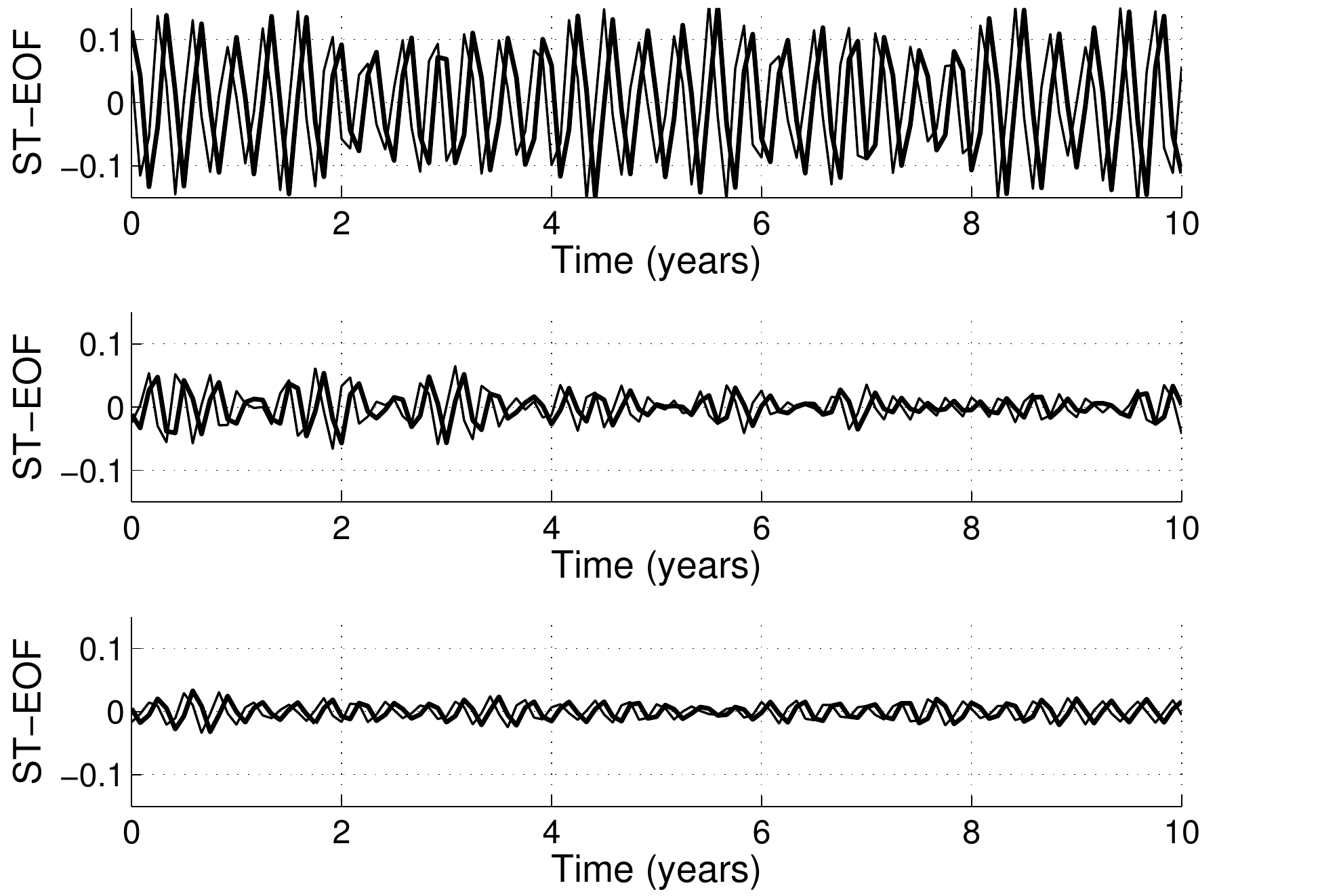}}
\subfigure[Pair 7-8]{\label{fig:MSSA_M121_STEOFs_Pair_7and8_Bath_NAO_MSL_MH}
\includegraphics[width=5.5cm]{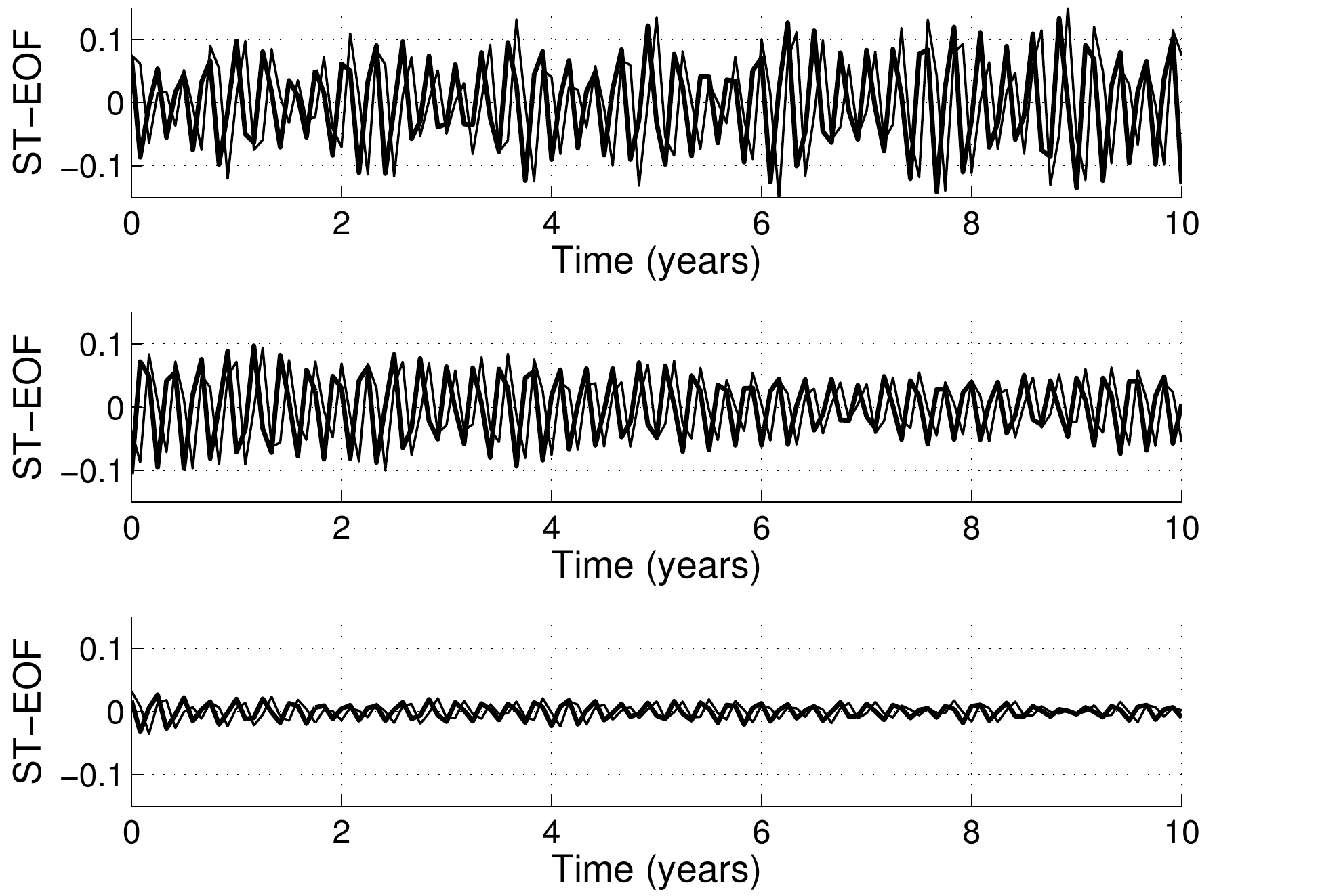}}\\
\subfigure[Pair 9-10]{\label{fig:MSSA_M121_STEOFs_Pair_9and10_Bath_NAO_MSL_MH}
\includegraphics[width=5.5cm]{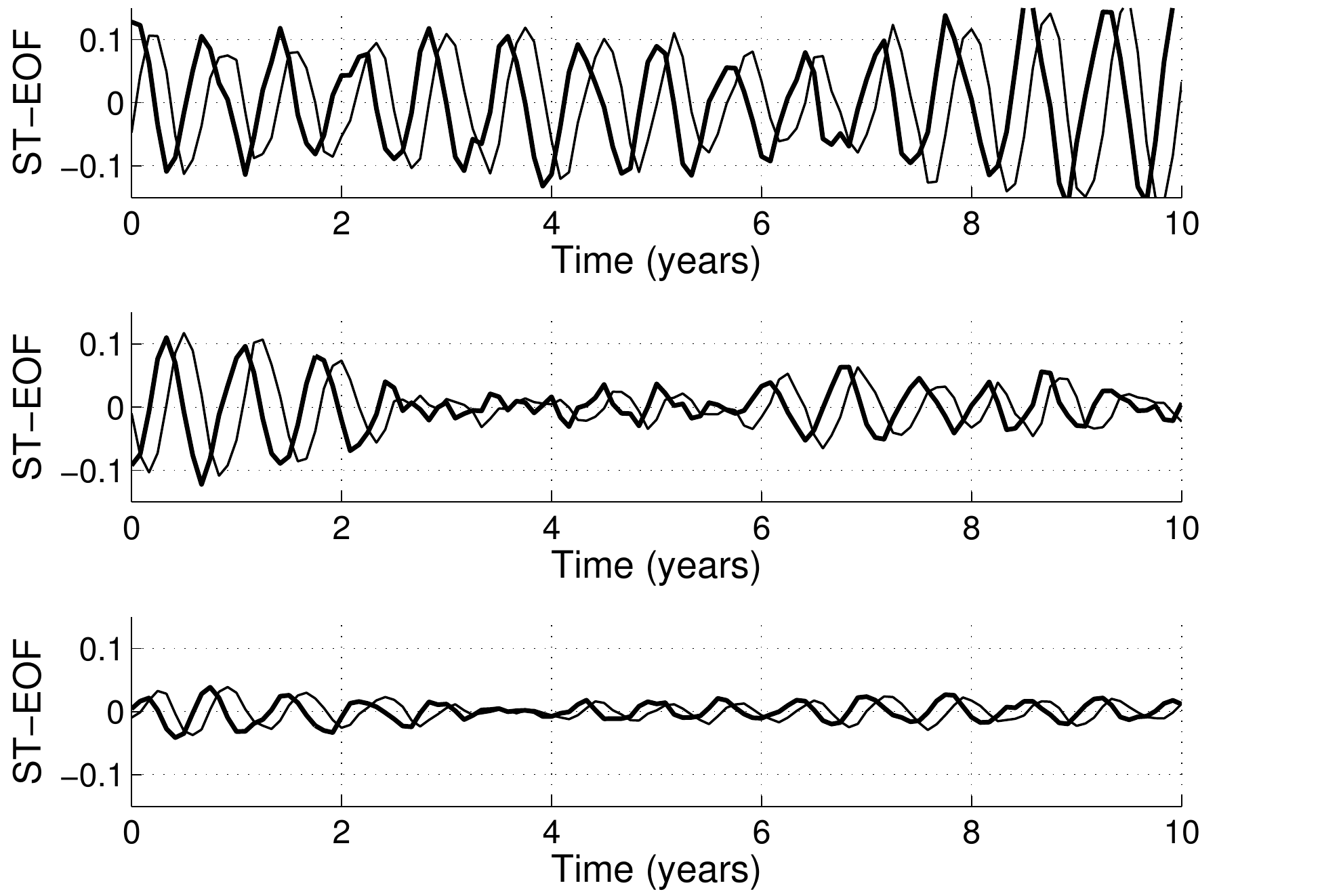}}
\subfigure[Pair 11-12]{\label{fig:MSSA_M121_STEOFs_Pair_11and12_Bath_NAO_MSL_MH}
\includegraphics[width=5.5cm]{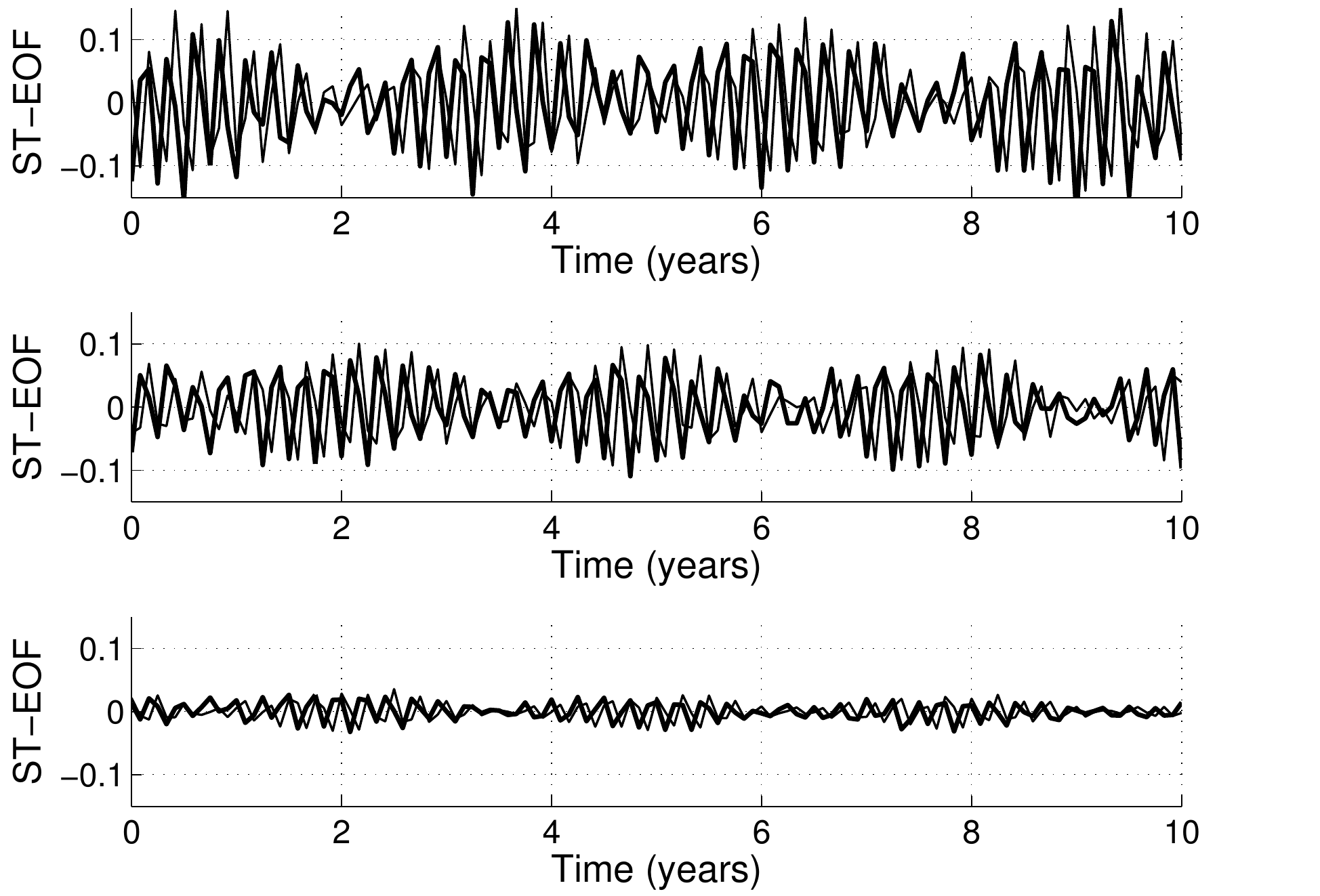}}\\
\caption{First six pairs of ST-EOFs of the MSSA components linked to potentially quasi-oscillatory pairs at all channels, with their associated resolved variance. The abscissa represents time (in months) and spans the window length $M=121$~months.}
\end{figure}

\begin{figure}[ht]
\includegraphics[width=12cm]{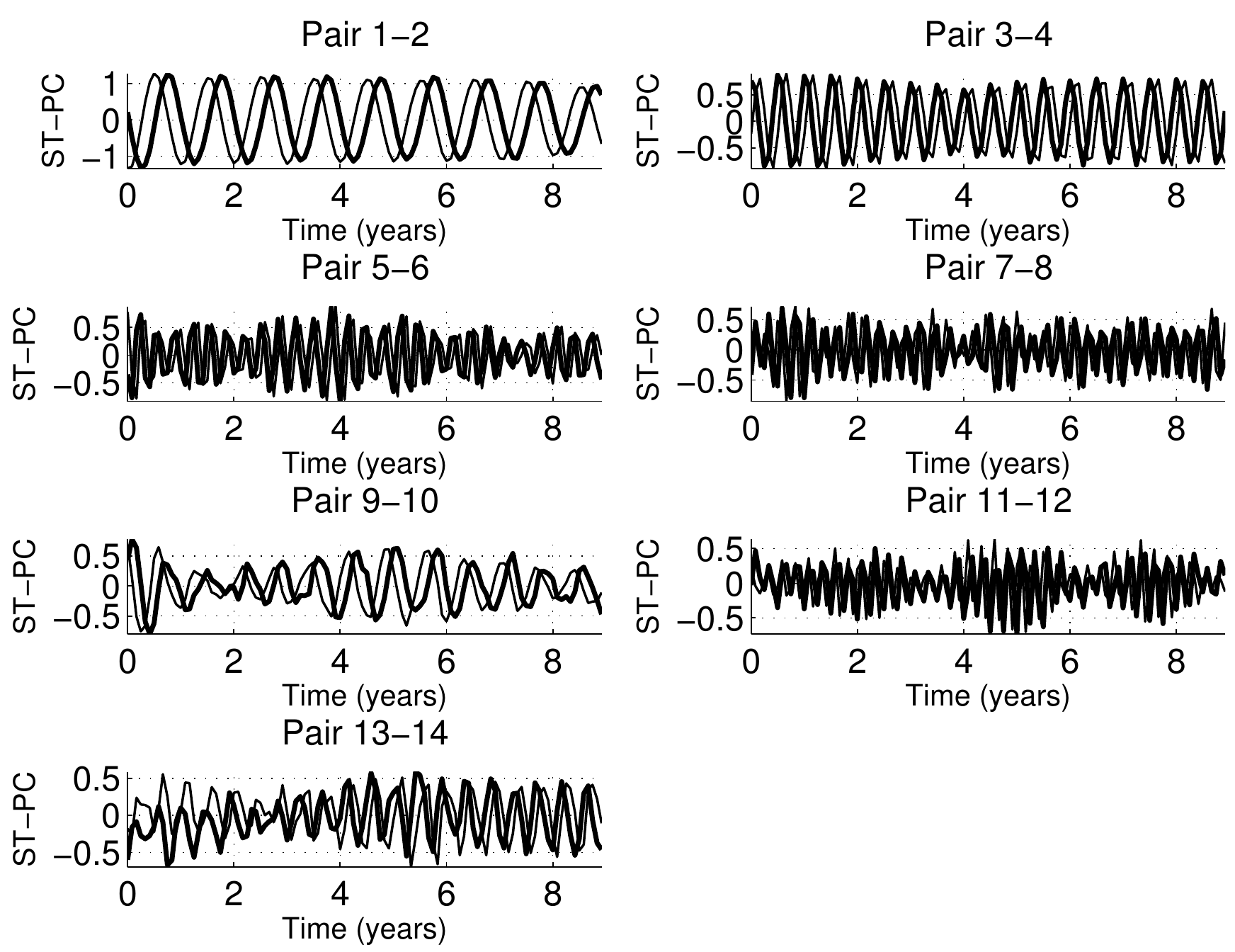}
\caption{\label{fig:MSSA_M121_PCsAtChannel1_Bath_NAO_MSL_MH} ST-PCs for consecutive MSSA components at channel~1; the ST-PCs span the window length $N'=108$~months.}
\end{figure}

\begin{figure}[ht]
\includegraphics[width=12cm]{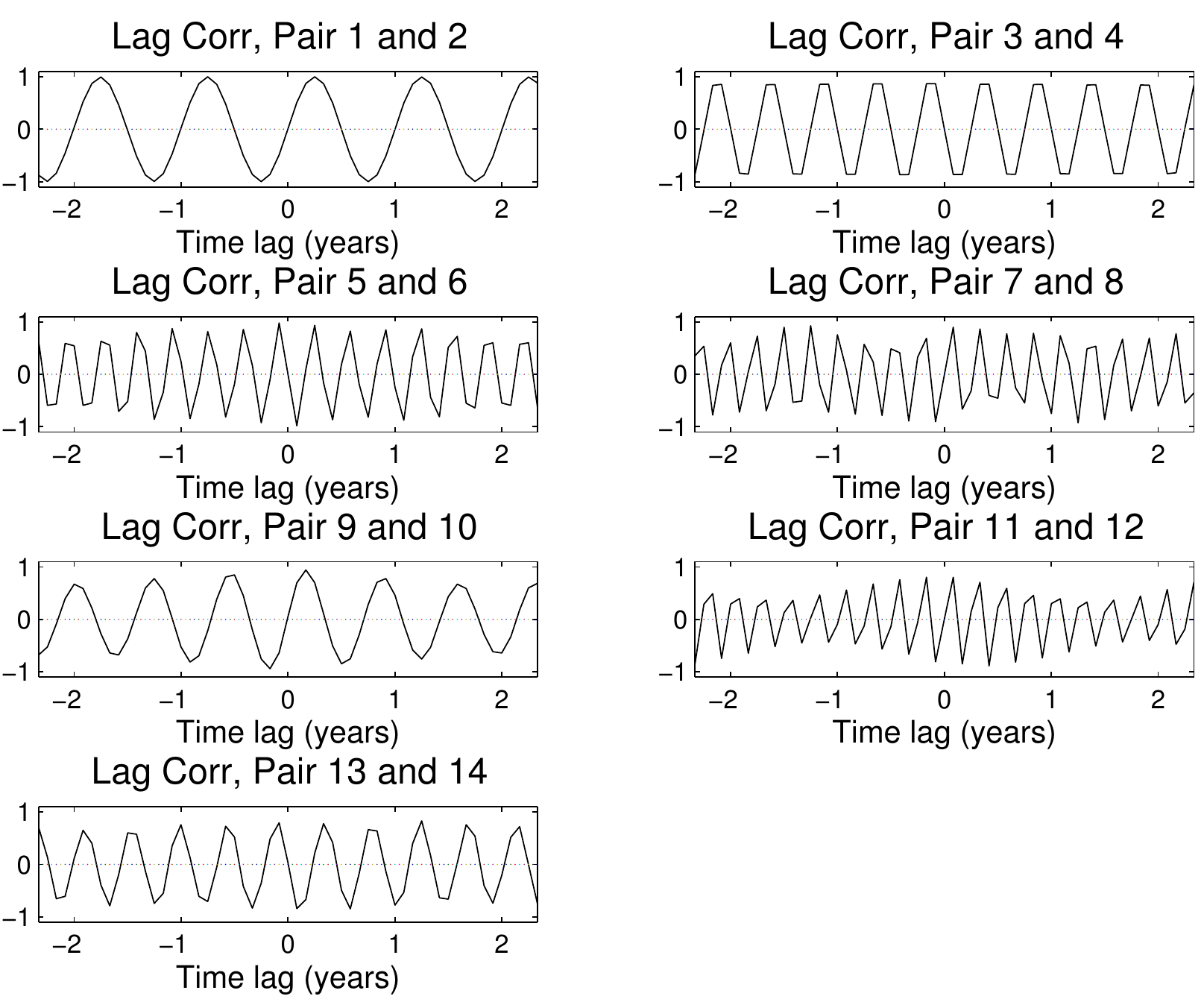}
\caption{\label{fig:MSSA_M121_PCPairs_LagCorr_Bath_NAO_MSL_MH} Lagged correlations for all ST-PC pairs.}
\end{figure}
\clearpage

\noindent {\large{\bf{Tables:}}}
\begin{table}[h]
\label{table_longshore}
\begin{center}
\begin{tabular}{|c|c|c|c|c|c|c|c|c|c|c|} \hline
TN & 190 & 188 & 62 & 58 \\ \hline
$\overline{y}$ & -91 & 0 & 1008 & 1097 \\ \hline
$\Delta y$ & 32 & 17 & 98 & 91 \\ \hline
\end{tabular}
\caption{Correspondence between transect number and transect longshore position for the four best surveyed transects at Duck, North Carolina, USA. $TN$ is the transect number, $\overline{y}$ the averaged longshore position $\Delta y$ the absolute error around this average.}
\end{center}
\end{table}

\begin{table}
\begin{center} {\small
\begin{tabular}{||c||c|c||c|c||} \hline
&\multicolumn{2}{|c||}{Transect 58} & \multicolumn{2}{|c||}{Transect 190} \\\hline \hline
x (m) & LPOs (yrs) & SPOs (yrs) & LPOs (yrs) & SPOs (yrs) \\ \hline \hline
120 &  $6.5(2)$     & $1(0.7)$  & $15.5(1.5)$  & $1(0.1)$,$2.6(0.1)$\\
130 &  $16.4(4)$  & $1(0.1)$  & $8(1)$ &  $1(0.1)$,$2.7(0.2)$\\
140 & $11.4(5.1)$    & $1(0.2)$ &  $1.8(0.2)$,$3.1(0.5)$  & $1(0.1)$,$2.4(0.2)$\\
    &     &  &  $5.6(0.4)$  & \\
150 & $3.1(0.6)$   & $1(0.1)$   & $6.5(0.5)$  & $2.1(0.2)$,$1(0.1)$\\
180 & $3.9(0.3)$  & $1(0.1)$  & $13(1)$, $1.7(0.1)$   & $2.4(0.2)$,$0.9(0.2)$\\
190 &  $4.7(0.8)$   & $1(0.1)$  & $1.7(0.3)$   & $2.2(0.2)$,$1(0.1)$\\
200 & $6.4(0.6)$   & all below 1  & $5.5(0.2)$  & $2.9(0.5)$,$1.4(0.2)$,$1(0.1)$\\
220 &  $7.9(2.4)$  & $0.8(0.1)$, $1.8 ( 0.1)$  & $3.6(0.4)$  & $4.4(0.6)$,$1.4(0.2)$,$1(0.1)$\\
230 &  $6.3(1.2)$   & $2.2(0.5)$  & $10(1)$  & $1.5(0.1)$,$1 ( 0.1)$\\
240 & $4.4(0.8)$   & $1 ( 0.1)$,$1.9 ( 0.2)$  & $12(1)$   & $2.9 ( 0.5)$\\
250 &  $5.4(0.9)$   & $1 ( 0.2)$,$1.8 ( 0.1)$  & $12(1)$  & $2.6(0.3)$,$1.6(0.3)$,$1(0.2)$ \\
260 &  $5.2(0.7)$  & $1.8 ( 0.2)$,$4.2 ( 0.6)$  & $14(2)$  & $2.5(0.3)$,$1.6(0.3)$,$1.2(0.2)$\\
310 & $5.8(0.7)$   & $1.1 ( 0.1)$,$1.7 ( 0.3)$  & $5.8 ( 0.2)$  & $1.2(0.2)$,$1.6(0.3)$,$2.5(0.3)$\\
340 &  $7.3(1.8)$  & $1.5(0.3)$  & $6.5(0.5)$  &  $3.1(0.4)$,$1.5(0.3)$,$1(0.2)$ \\
350 &  $9.8(4)$  & $2.1(0.9)$  & $6.5(0.5)$,$8(0.4)$  & $1(0.2)$,$1.5(0.2)$,$2.2 ( 0.1)$\\
370 & $3.4(0.5)$  & $1.4 ( 0.1)$, $4.8 ( 1.1)$  & $6.2(0.4)$,$9.3(0.6)$  & $1(0.5)$,$2.2(0.2)$\\
390 & $3.4(0.2)$     & $2.2(0.6)$  & $5.8(0.4)$  & $1(0.5)$,$1.6(0.5)$,$2.3(0.2)$\\
410 & $3.3 (0.9)$  & $1.4 (0.1)$  & $4.3 (0.2)$  & $1(0.1)$,$2(0.2)$,$3.2 (0.6)$\\
470 & $6.7(0.8)$   & $2.5(0.9)$  & $3.2(0.2)$,$4.8(0.2)$  & same as 410\\
480 &  $10.3(6.2)$   & $2.6(0.6)$,$3.3(0.8)$ & $3.3(0.2)$  & same as 410 \\
490 & $7.1(2)$  & $2.5(0.5)$  & $4.8(0.2)$  & $1(0.1)$,$1.3(0.4)$,$3.4(0.6)$\\ \hline
\end{tabular}  }
\end{center}
\caption{\label{table_peaks} Periods of dominant oscillatory patterns at selected locations along transects~$58$ and~$190$, with the uncertainty or confidence interval in brackets.}
\end{table}

\begin{table}
\label{table_Ucomponents}
\begin{center} {\small
\begin{tabular}{|c|c|c|c|c|} \hline
\multicolumn{5}{|c|}{EOF values with all seabed surveys}\\ \hline
Forcing: & EOF1      & EOF2      & EOF3     & EOF4     \\ \hline
NAO      & $0.998$  &-$0.067$  &-$0.012$ &-$0.0005$ \\
MWH      & $0.067$  & $0.998$  & $0.0005$&-$0.0009$ \\
MWL      &-$0.012$  & $0.001$  &-$0.999$ & $0.044$  \\ \hline
\multicolumn{5}{|c|}{EOF values with Transects 58 and 190 only}\\ \hline
Forcing: & EOF1      & EOF2      & EOF3     & EOF4      \\ \hline
NAO      &$0.998$  & -$0.067$  & -$0.013$ &-$0.0001$ \\
MWH      &$0.067$  &$0.998$  &$0.0004$& $0.003$  \\
MWL      & -$0.012$  &$0.001$  & -$0.940$ &-$0.335$  \\ \hline
\end{tabular}  }
\end{center}
\caption{Table of components of V associated with the first 4 main SVD components, for the 3 potential forcings considered: the North Atlantic Oscillation (NAO), Monthly-averaged Wave Heights (MWH), and the monthly mean water level (MWL). The forcings have been arranged so that the largest elements in the Forcings/EOFs matrix are in the diagonal.}
\end{table}

\end{document}